\documentclass[twoside]{siamltex}

\catcode`\@=11
\def\ltsim{\mathrel{\vcenter{\m@th\offinterlineskip
\hbox{$\hfill<\hfill$}\kern.5ex\hbox{$\hfill\sim\hfill$}}}}
\catcode`\@=12
\catcode`\@=12

\usepackage{graphicx}

\begin{document}
%\draft %prints PACS numbers in

\title{Depinning transitions in discrete reaction-diffusion equations
\thanks{This research was supported by the Spanish MCyT grant
BFM2002-04127-C02, by the Third Regional Research Program of the Autonomous
Region of Madrid (Strategic Groups Action), and by the European Union under
grant RTN2-2001-00349.  Received by the editors of SIAM J. Appl. Math.
on  31 May 2001. Manuscript number 0390061.}}
\author{A. CARPIO\footnotemark[2]
\and L. L. BONILLA\footnotemark[3] \footnotemark[4]}
\date{\today}
\maketitle

\renewcommand{\thefootnote}{\fnsymbol{footnote}}

\footnotetext[2]{Departamento de Matem\'atica Aplicada, Universidad
Complutense de Madrid, 28040 Madrid, Spain. }

\footnotetext[3]{Departamento de Matem\'aticas, Escuela Polit{\'e}cnica
Superior, Universidad Carlos III
de Madrid, Avenida de la Universidad 30, 28911 Legan{\'e}s, Spain.
}
\footnotetext[4]{Also: Unidad Asociada al
Instituto de Ciencia de Materiales de Madrid (CSIC).}

\renewcommand{\thefootnote}{\arabic{footnote}}
\renewcommand{\theequation}{\arabic{section}.\arabic{equation}}
\newcommand{\fin}{\newline \rule{2mm}{2mm}}
\def\RR{\hbox{{\rm I}\kern-.2em\hbox{\rm R}}}
\def\pRR{\hbox{{\tiny \rm I}\kern-.1em\hbox{{\tiny \rm R}}}}
\def\ZZ{\hbox{{\rm Z}\kern-.42em\hbox{\rm Z}}}

\begin{abstract}
We consider spatially discrete bistable reaction-diffusion equations that
admit wave front solutions. Depending on the parameters involved, such
wave fronts appear to be pinned or to glide at a certain speed. We study
the transition of traveling waves to steady solutions near threshold and
give conditions for front pinning (propagation failure). The critical
parameter values are characterized at the depinning transition and an
approximation for the front speed just beyond threshold is given.
\end{abstract}

%\begin{multicols}{2}
%\narrowtext


%\begin{AMS}
34E15, 92C30. \hspace{2cm} Date: \today
%\end{AMS}

\pagestyle{myheadings}
\thispagestyle{plain}
\markboth{A. CARPIO AND L. L. BONILLA} {Depinning in discrete
equations}

%%%%%%%%%%%%%%%%%%%%%%%%%%%%%%%%%%%%%%%%%%%
%\newpage
%\baselineskip 4.6ex
%\parskip 2.3ex
%
\setcounter{equation}{0}
\section{Introduction}
\label{sec:introduction}

Spatially discrete systems describe physical reality in many different
fields: atoms adsorbed on a periodic substrate \cite{cha95}, motion of
dislocations in crystals \cite{nab67}, propagation of cracks in a brittle
material \cite{sle81}, microscopic theories of friction between solid bodies
\cite{ger01}, propagation of nerve impulses along myelinated fibers
\cite{kee87,kee98}, pulse propagation through cardiac cells \cite{kee98},
calcium release waves in living cells \cite{bug97}, sliding of charge density
waves \cite{cdw}, superconductor Josephson array junctions \cite{jj}, or
weakly coupled semiconductor superlattices \cite{bon94,car00}. No one really
knows why, but spatially discrete systems of equations often times have
smooth solutions of the form $u_n(t)= u(n-ct)$, which are monotone functions
approaching two different constants as $(n-ct)\to\pm\infty$. Existence of
such {\em wave front} solutions has been proved for particular discrete
systems having dissipative dynamics \cite{zin92}. In the case of discrete
systems with conservative dynamics, a wave front solution was explicitly
constructed by Flach et al.\ \cite{fla99}. A general proof of wave front
existence for discrete conservative systems with bistable sources is
lacking.  

A distinctive feature of spatially discrete reaction-diffusion systems (not
shared by continuous ones) is the phenomenon of wave front pinning:
for values of a control parameter in a certain interval, wave fronts joining
two different constant states fail to propagate \cite{kee98}.
When the control parameter surpasses a threshold, the wave front depins and
starts moving \cite{kee87,cdw,nab67,car00}. The existence of such thresholds
is thought to be an intrinsically discrete fact, which is lost in continuum
aproximations. The characterization of propagation failure and front
depinning in discrete systems is thus an important problem, which is not yet
well understood despite the numerous inroads made in the literature
\cite{kee87,bug97,cdw,nab67,kev00,kev01,kla00,mit98,kin01,spe94,spe97,spe99}.

In this paper, we study front depinning for infinite one-dimensional
nonlinear spatially discrete reaction-diffusion (RD) systems. When
confronted with a spatially discrete RD system, a possible strategy is to
approximate it by a continuous RD system. For generic nonlinearities, the
width of the pinning interval is exponentially small as the continuum limit
is approached. Pinning in the continuum limit has been analyzed by many
authors using exponential asymptotics, also known as asymptotic beyond all
orders. As far as we can tell, usage of these techniques for discrete 
equations goes back to two classic papers by V.L. Indenbom \cite{ind58} (for
the FK potential) and by J.W. Cahn \cite{cah60} (for the double-well
potential). In both cases, an exponential formula for the critical field was
derived by means of the Poisson sum rule. In the context of dislocation
motion, exponential formulas for the depinning shear stress of the
Peierls-Nabarro (PN) model were found earlier by Peierls \cite{pei40} and
Nabarro \cite{nab57}. Descriptions of wave front pinning near the continuum
limit can also be found in more recent work \cite{hak93,kev00,kin01}. 

Analyzing the continuum limit of a discrete system by means of exponential
asymptotics is a costly strategy to describe pinning for two reasons. It
is not numerically accurate as we move away from the continuum limit, and it
ceases to be useful if convective terms \cite{abc} or disorder \cite{car02}
alter the structure of the discrete system (quite common in applications).
Thus other authors have tried to describe the opposite strongly discrete
limit. For discrete RD equations, Erneux and Nicolis \cite{ern93} studied a
finite discrete RD equation with a cubic nonlinearity, Dirichlet boundary
condition at one end and a Neumann boundary condition at the other end. They
considered a particular limit in which two of the three zeroes of the cubic
nonlinearity coalesced as diffusivity went to zero. Erneux and Nicolis's
calculation is essentially a particular case of our active point
approximation that involves  only one active point and makes an additional
assumption on the nonlinearity (not needed in our calculations). They found
that the wave front velocity scales as the square root of $(d-d_c)$ ($d$ is
the diffusivity and $d_c$ its critical value at which wave fronts are
pinned). Essentially the same results can be found in the Appendix of Ref.\
\cite{kin01}. Kladko et al \cite{kla00} introduced an approximation called
the single active site theory. In this approximation, the wave front is
described by two linear tails (solution of the RD equation linearized about
each of the two constants joined by the front) {\em patched} at one point.
This approximation was used to estimate the critical field for wave front
depinning.

By a combination of numerical and asymptotic calculations, we arrive at the
following description \cite{CB,abc}. The nature of the depinning transition
depends on the nonlinearity of the model, and is best understood as
propagation failure of the traveling front. Usually, but not always, the
wave front profiles become less smooth as a parameter $F$ (external field)
decreases. They become {\em discontinuous} at a critical value $F_c$. Below
$F_c$, the front is pinned at discrete positions corresponding to a stable
steady state. As a consequence of the maximum principle for spatially
discretized parabolic equations, stationary and moving wave fronts cannot
simultaneously exist for the same value of $F$ \cite{car99}. This is {\em
not} the case for chains with conservative dynamics, that are spatially
discretized hyperbolic equations without a maximum principle. For chains 
with conservative Hamiltonian dynamics, an inverse method due to Flach,
Zolotaryuk and Kladko \cite{fla99} explicitly shows that stationary and
moving wave fronts may coexist for the same value of the parameters.

\begin{figure}
\begin{center}
\includegraphics[width=8cm]{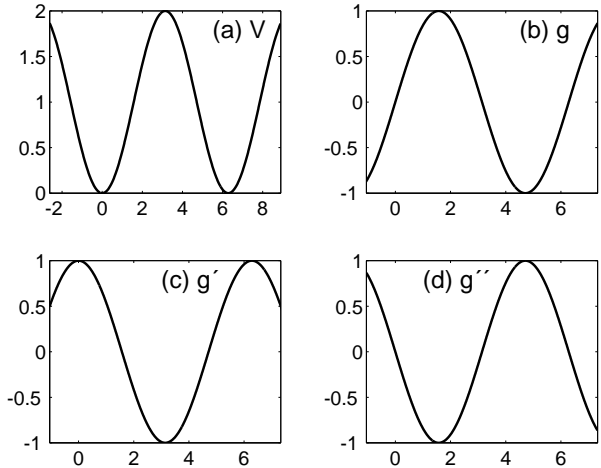}
\caption{FK model: (a) Potential $1-cos(x)$, (b) Source term
$g(u)=\sin(u)$, (c) $g'(u)=\cos(u)$, (d) $g''(u)=-\sin(u)$.}
\label{fig0}
\end{center}
\end{figure}

We consider chains of diffusively coupled overdamped oscillators in a 
potential $V$, subject to a constant external force $F$:
\begin{eqnarray}
{du_{n}\over dt} = u_{n+1}-2u_n + u_{n-1} + F -
A\,  g(u_n).   \label{Fd}
\end{eqnarray}
Here $g(u)=V'(u)$ is at least $C^1$ and it presents a `cubic' nonlinearity
(see Fig. \ref{fig0}), such that $A\, g(u)-F$ has three zeros, $U_1(F/A)
<U_2(F/A) <U_3(F/A)$ in a certain force interval ($g'(U_i(F/A)) >0$ for
$i=1,3$, $g'(U_2(F/A))<0$). Provided $g(u)$ is odd with respect to $U_2(0)$,
there is a symmetric interval $|F|\leq F_c$ where the discrete wave fronts
joining the stable zeros $U_1(F/A)$ and $U_3(F/A)$ are pinned
\cite{kee87,car99}. For $|F|>F_c$, there are smooth traveling wave fronts,
$u_n(t)=u(n-ct)$, with $u(-\infty)= U_1$ and $u(\infty)=U_3$, as proved in 
\cite{zin92,car99}. The velocity $c(A,F)$ depends on $A$ and $F$ and it
satisfies $cF<0$ and $c\to 0$ as $|F|\to F_c$ \cite{car99}. Examples are the
overdamped Frenkel-Kontorova (FK) model ($g=\sin u$; see Fig.\ref{fig0})
\cite{FK} and the quartic double well potential ($V=(u^2-1)^2 /4$). Less
symmetric nonlinearities yield a non-symmetric pinning interval and our
analysis applies to them with trivial modifications. Note that coexistence
of fronts traveling in opposite directions can occur in the case of 
conservative systems, but not for Eq.\ (\ref{Fd}) due to the maximum 
principle (which is the basis of comparison techniques) \cite{car99}.

\begin{figure}
\begin{center}
\includegraphics[width=10cm]{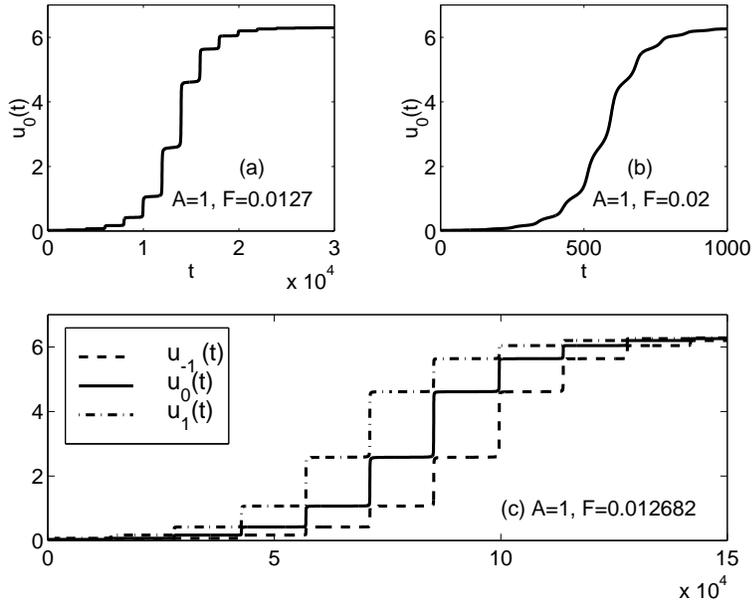}
\caption{Wave front profiles for the overdamped FK model when $A=1$ near
$F_c$.}
\label{fig1}
\end{center}
\end{figure}

For the overdamped FK model given by Eq.\ (\ref{Fd}) with $g=\sin u$, Fig.\
\ref{fig1} shows wave front profiles near the critical field. Individual
points undergo abrupt jumps at particular times, which gives the misleading
impression that the motion of the discrete fronts proceeds by successive
jumps. Actually, the points remain very close to their stationary values at
$F=F_c$, say $u_n(A,F_c)$, during a very long time interval of order
$|F-F_c|^{-{1\over 2}}$. Then, at a specific time, {\em all} the points
$u_n(t)$ jump to a vicinity of $u_{n+1}(A,F_c)$. The method of matched
asymptotic expansions can be used to describe this two-stage motion of the
points $u_n(t)$. Then the wave front profile can be reconstructed by using
the definition $u_n(t)= u(n-ct)$. The slow stage of front motion is
described by the normal form of a saddle-node bifurcation and it yields an
approximation to the wave front velocity, that scales with the field as
$|F-F_c|^{{1\over 2}}$. This scaling has been mentioned by other authors: it
was found numerically in \cite{ama01} and by means of exponential
asymptotics in the limit $A$ small in \cite{kin01}. It is also conjectured
in \cite{kev01} on the correct basis that the depinning transition consists
of a saddle-node bifurcation (a similar claim was stated in \cite{mit98} for
continuous reaction diffusion equations with localized sources). However,
the derivation of the {\em local} saddle-node normal form and the correct
description of the {\em global} saddle-node bifurcation involving matching
with a fast stage during which the front jumps abruptly one lattice period
were apparently omitted by the authors of Ref. \cite{kev01}, who used energy
arguments. Our picture of the wave front depinning transition has
essentially been corroborated by King and Chapman in the continuum limit (as
an appropriate dimensionless lattice length goes to zero) who used
asymptotics beyond all orders \cite{kin01}. An independent confirmation
follows from F\'ath's calculations for a spatially discrete
reaction-diffusion equation with a piecewise linear source term \cite{fat98}
(except that the velocity should scale differently with $|F-F_c|$ in this
case).

For exceptional nonlinearities, the wave front does not lose continuity as
the field decreases. In this case, there is a continuous transition between
wave fronts moving to the left for $F>0$ and moving to the right for $F<0$:
as for continuous systems, front pinning occurs only at a single field value
$F=0$ \cite{kin01,fla99,spe94,spe97,spe99}. Wave front velocity scales then
linearly with the field. We discuss the characterization of the critical
field (including analytical formulas in the strongly discrete limit),
describe depinning anomalies (discrete systems having zero critical field
\cite{spe94,spe97,spe99,fla99}), and give a precise characterization of
stationary and moving fronts near depinning (including front velocity) by
singular perturbation methods. Our approximations show excellent agreement
with numerical simulations. 

The rest of the paper is organized as follows. In section \ref{sec:pinning},
we characterize wave front depinning. We also explain that pinning of wave
fronts normally occurs at force values belonging to an interval with nonzero
length. However, there are nonlinearities for which pinning occurs only at
$F=F_c=0$. In Section \ref{sec:asymptotic}, we present a theory of wave
front depinning for the strongly discrete case ($A$ large). This theory
enables us to predict the critical field and the speed and shape of the wave
fronts near threshold. The main ideas of our theory are very simple. First
of all, a wave front profile $u_n(t)= u(n-ct)$ can be reconstructed if we
follow the motion of one point during a sufficiently long time interval.
Secondly, the analysis of Eq.\ (\ref{Fd}) is complicated by the presence of
the discrete diffusion term, $u_{n+1}-2u_n + u_{n-1}$. Previous authors have
tried to approximate this term by its continuum limit (corresponding to
$A\to 0$), which leads to using exponential asymptotics \cite{kin01} (see
also \cite{kev00} on using exponential asymptotics for the Hamiltonian
version of our model).  However, we are only interested in constructing
solutions of Eq.\ (\ref{Fd}) joining constant values. For sufficiently large
$A$ (say $A=0.1$ for the FK model), $u_i$ is approximately either $U_1(F/A)$
or $U_3(F/A)$ except for a finite number of points (the {\em active} points).
Then we can approximate the infinite system (\ref{Fd}) by a closed system of
ordinary differential equations (only one equation for $A\geq 10$ in the FK
model). The depinning transition is a global bifurcation of this system, as
explained in Section \ref{sec:asymptotic}. Some auxiliary technical results
are collected in Appendix A.

\setcounter{equation}{0}
\section{Front pinning as propagation failure}
\label{sec:pinning}
To describe monotone stationary solutions of (\ref{Fd}) joining
$U_1(F/A)$ and $U_3(F/A)$ for $|F|\leq F_c$, it is better to start by
considering traveling wave fronts for $|F|> F_c$. It has been proved (and
corroborated by numerical calculations) that traveling wave fronts
and stationary profiles cannot coexist at the same value of $F$
\cite{car00}. Furthermore, numerical computations of wave fronts near
the critical fields $F_c$ for the FK and other usual potentials show
staircase-like wave front profiles, which sharpen as $F$ approaches
$F_c$. At $F=F_c$, a series of gaps open up and one is left with a
discontinuous stationary profile $s(x)$ solving
\begin{eqnarray}
s(x+1)-2s(x)+s(x-1)=Ag(s(x))-F_c, \quad x\in \bf \RR, \nonumber \\
s(-\infty)=U_1(F_c/A),\quad s(\infty)=U_3(F_c/A). \label{e1}
\end{eqnarray}
The profile $s(x)$ is increasing and piecewise constant. The sequence of
constant values attained by $s(x)$ defines a steady solution $u_n$ of
(\ref{Fd}) with $F=F_c$. A stationary solution can thus be understood
as a wave front that fails to propagate and is {\em pinned} at discrete
values.  Fig.\ \ref{fig1} illustrates the pinning transition for the FK
model with $A=1$. As $F$ decreases from $0.02$ to $0.0127$, a series of
steps are formed. Fig.\ \ref{fig1} (c) depicts the paths described
by three consecutive points. All profiles look identical and are obtained
by shifting anyone of them some multiple of a certain constant length.
This implies that the length of all steps in the profile is the same and
all the points $u_n(t)$ in (\ref{Fd}) proceed to climb the next step in
the staircase at the same time. This behavior indicates that the
wave front is a  traveling wave, $u_n(t)=u(n-ct)$. Proofs of this fact
for some sources can be found in Ref.\ \cite{zin92}.

\subsection{Limiting front profile at the critical field}
Let us start by showing that the limit of the traveling waves as $F\to
F_c$ is singular if $F_c>0$. This fact can be guessed from the
differential-difference equations satisfied by the wave profiles. The
traveling waves for $|F|>F_c$ have the form $u_n(t)=u(n-ct)$, where the
profile $u(z)$ solves
\cite{car99}
\begin{eqnarray}
-c u_z=u(z+1)-2u(z)+u(z-1)-Ag(u(z))+F \quad z\in \bf \RR, \nonumber \\
u(-\infty)=U_1(F/A),\quad u(\infty)=U_3(F/A). \label{e2}
\end{eqnarray}
The solution $u$ is as smooth as allowed by $g(u)$ ($u$ is $C^{k+1}$ if
$g(u)$ is $C^k$, with $k\geq 1$). Mutiplying then Eq.\ (\ref{e2}) by $u_z$
and integrating it, we get
\begin{eqnarray}
-c \int_{-\infty}^{\infty} u_z^2 dz = F\,
\left[U_3\left({F\over A} \right) - U_1\left({F\over A} \right)\right].
\label{e3}
\end{eqnarray}
A first obvious conclusion is that the sign of $c$ is opposite to the
sign of $F$. Let $F_c$ be positive. As $F\to F_c$, $c\to 0$ and $F\,
[U_3(F/A) -U_1(F/A)]\rightarrow F_c\, [U_3(F_c/A)-U_1(F_c/A)] \neq 0$.
Therefore the integrals $\int u_z^2 dz\to\infty$ as $F\to F_c$.
Thus the limiting profile must be discontinuous if $F_c> 0$.

If $F_c=0$, the relation (\ref{e3}) can be used to show that Eq.\ (\ref{e1})
has a smooth solution. In fact, provided $c\sim -K\, F$ ($K>0$) as $F\to 0$,
we can use (\ref{e3}) to bound uniformly the derivatives of the solutions $u$
in (\ref{e2}) for $F\neq 0$. Then we obtain a smooth solution of Eq.\
(\ref{e1}) in the limit as $F\to 0$. We will come back to this question later
on in Subsection 2.3. Note that the stationary equation $s(x+1) - 2s(x) +
s(x-1) = Ag(s)-F$ has no continuous solutions joining $U_1(F/A)$ to
$U_3(F/A)$ unless $F=0$. To see this \cite{car99}, we multiply the equation
by $s_x$ (in the sense of distributions if necessary) and integrate to get
$F=0$.

\subsection{Characterization of the critical field}
Some results are available in the continuum limit $A \rightarrow 0$. For
$g=\sin u$, it is well known that $F_c$ vanishes exponentially fast as $A$
goes to zero.  An exponential formula for $F_c$ was first found by Indenbom
\cite{ind58} using the Poisson sum rule (following the calculations of the
PN energy barrier for the PN model by Peierls \cite{pei40} and Nabarro
\cite{nab57}), and numerically checked by Hobart \cite{hob65} in the context
of the Peierls stress and energy for dislocations. For the discrete bistable
RD equation, Cahn \cite{cah60} derived an exponential dependence of $F_c$ by
a similar technique. Related ideas can be found in Kladko et al
\cite{kla00}. These arguments can be used for other potentials and suggest
that $F_c\sim C\, e^{-\eta/\sqrt{A}}$ as $A\to 0+$ (with positive $C$ and
$\eta$ independent  of $A$) holds for a large class of nonlinearities. Using
exponential asymptotics, King and Chapman \cite{kin01} have obtained precise
formulas for the critical field and the wave front velocity of a discrete RD
equation. Particularized to the FK potential, their formulas for the
critical field and for the wave front velocity after depinning are $F_c \sim
\Lambda\, e^{-\pi^2/[2 \sinh^{-1} (\sqrt{A}/2)]}$, $\Lambda\approx 356.1$,
and $c\sim D\,\sqrt{(F^2- F_c^2)/A}$, respectively. This later result agrees
with the scaling law $c\sim |F-F_c|^{ {1\over 2}}$, found in a large class
of discrete RD equations \cite{car00,CB,abc,kev01} and in continuous
equations with localized sources \cite{mit98}. However, exponential
asymptotics \cite{kin01} does not work for $A$ large. We shall therefore
follow a different approach. We shall begin by considering stationary
increasing discrete front profiles and study under which conditions they
start moving. Since stationary fronts are pinned wave fronts, we can call
the transition from stationary to moving fronts the {\em depinning
transition}.

Two facts distinguish the depinning transition: (i) the smallest eigenvalue
of (\ref{Fd}) linearized about a stable stationary profile becomes zero (see
below), and (ii) stationary and moving wave fronts cannot coexist for the
same values of the field. First of all, the following comparison principle
\cite{kee87} for (\ref{Fd}) can be used to show that stationary and
traveling wave fronts cannot coexist for the same value of $F$ \cite{car99}:

{\it {\bf Comparison principle:} Assume that we have two configurations
$w_n(t)$ and $l_n(t)$. If initially  $w_n(0) \geq l_n(0)$, $\forall n$
and at any later time $t>0$
\begin{eqnarray}
{dw_{n}\over dt} \geq w_{n+1}-2w_n + w_{n-1} - A g(w_n) + F,
\label{sup} \\
{dl_{n}\over dt} \leq l_{n+1}-2l_n + l_{n-1} - A g(l_n) + F,
\label{sub}
\end{eqnarray}
for all $n\in \ZZ$, then necessarily $w_n(t)\geq l_n(t)$ for all $n$ and
$t$. $w_n$ satisfying (\ref{sup}) is said to be a supersolution. $l_n$
satisfying (\ref{sub}) is said to be a subsolution.}

Front pinning can be proved using stationary sub and supersolutions,
which can be constructed provided the stationary solution is linearly
stable. The smallest eigenvalue of the linearization of (\ref{Fd}) about
a stationary profile $u_n(A,F)$, $u_n(t)= u_n(A,F) + v_n e^{-\lambda
t}$, is given by
\begin{eqnarray}
\lambda_1(A,F)=\mbox{Min}\, {\sum [(v_{n+1}-v_n)^2 +A g'(u_n(A,F)) v_n^2]
\over \sum v_n^2}\,, \label{var}
\end{eqnarray}
over a set of functions $v_n$ which decay exponentially as $n\to\pm
\infty$. We show in the Appendix that the minimum is attained at a positive
eigenfunction. 

\begin{figure}
\begin{center}
\includegraphics[width=6cm]{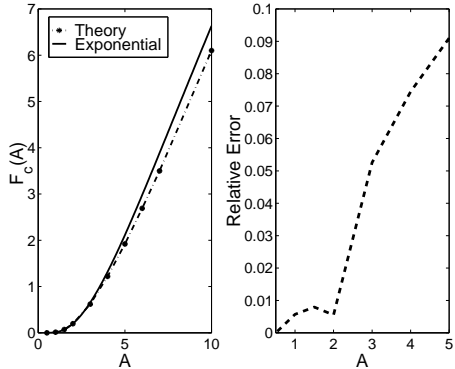}
\caption{(a) Critical field, $F_c(A)$, for $A\in \left({1\over 2}, 10
\right)$. We have compared the result of approximately solving
$\lambda_1(A,F)=0$ for $F$ as a function of $A$ (see Appendix A) to the
asymptotic result $F_c \sim  356.1\, e^{-\pi^2/[2 \sinh^{-1}(\sqrt{A}/2)]}$
of Ref.\ \cite{kin01}. (b) Relative error of the exponential asymptotics
approximation. }
\label{fig10}
\end{center}
\end{figure}

The critical field can be uniquely characterized by $\lambda_1(A,F_c) = 0$
and $\lambda_1(A,F)>0$ for $|F| <F_c$. The details are given in Appendix A.
Notice that $\lambda_1(A,F)>0$ implies that Eq.\ (\ref{e1}) does not have
smooth solutions $s(x)$. For otherwise, $v_n= s'(n)$ is an eigenfunction
corresponding to $\lambda_1=0$ as it happens in the continuum limit. The
previous characterization is the base of a procedure to calculate $F_c(A)$.
In Section \ref{sec:asymptotic}, we shall show that wave fronts near the
depinning transition are described by a reduced system of equations for a
finite number of points $u_n(t)$ which ``jump'' from about a discrete value
corresponding to the stationary solution, $u_n(A,F_c)$, to the next one,
$u_{n+1}(A,F_c)$, during front motion. The smallest eigenvalue for the
linearization of the reduced system of equations about a stationary solution
approximates well $\lambda_1$. The critical field obtained by this procedure
has been depicted in Fig.\ \ref{fig10} for the FK potential and compared to
King and Chapman's asymptotic result (obtained by keeping two terms in their
formulas). Notice that the asymptotic result loses accuracy as $A$
increases. 

\begin{figure}
\begin{center}
\includegraphics[width=6cm]{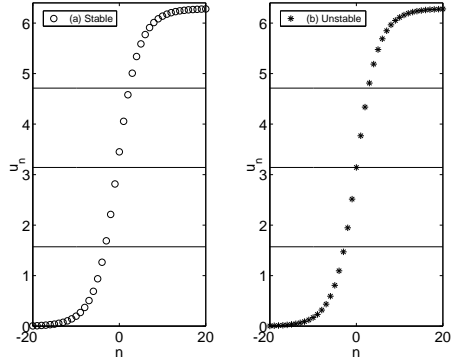}
\caption{Stable and unstable dislocations for the FK model when $F=0$ and
$A=0.1$.}
\label{fig2}
\end{center}
\end{figure}

\begin{figure}
\begin{center}
\includegraphics[width=10cm]{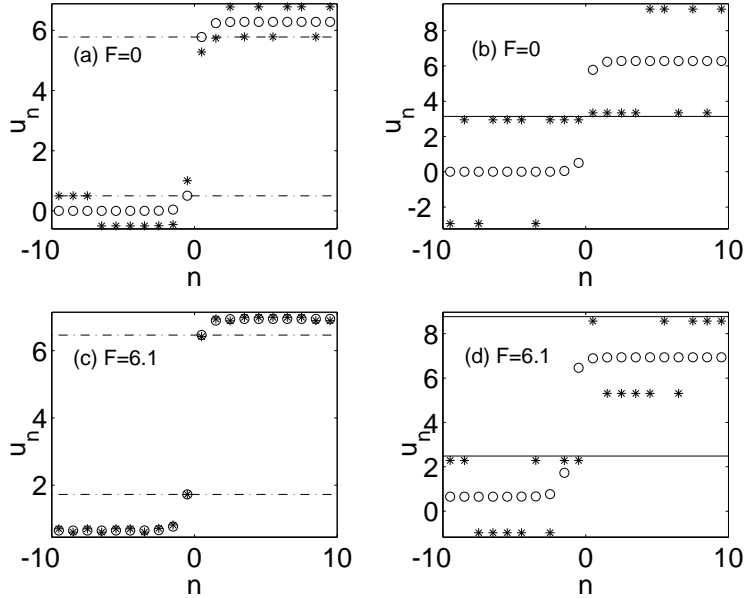}
\caption{Initial condition $u_n(0)$ (asterisks) and its large time limit, the
stable dislocation (circles), for the FK model with $A=10$. $F=0$ for (a) and
(b) and $F=6.1<F_c$ for (c) and (d). The initial points are selected as
indicated in the text.}
\label{fig12}
\end{center}
\end{figure}

Equation (\ref{var}) shows that the critical field is positive for large
$A$ and typical nonlinearities. In fact, consider the FK potential. For
$F=0$ there are two one-parameter families of stationary solutions which are
symmetric with respect to $U_2$ (see Fig. \ref{fig2}), one taking on the
value $U_2$ (unstable dislocation), and the other one having $u_n \neq U_2$
(stable dislocation) \cite{hob65,car99}. The centers of two stable (or two
unstable) dislocations differ in an integer number of lattice periods.
Except for a possible rigid shift, the stable dislocation, $u_n(A,0)$, is a
dynamically stable stationary solution towards which step-like initial
conditions evolve. Figs. \ref{fig12} (a) and (b) show two initial conditions
that evolve (exponentially fast) towards the stable dislocation. Half the
initial points, $u_n(0)$, have been selected to be below $U_2$, and the
other half are above this value. In Fig. \ref{fig12}(a), $u_n(0)-u_{n}(A,0) =
\epsilon_n$, where $\epsilon_n$ are real random numbers with $|\epsilon_n|<
0.5$. In Fig. \ref{fig12}(b), $u_n(0)-U_{1,3} = \delta_n B$, $0<B=U_2-U_1 -
0.2$, and $\delta_n$ takes randomly on the values 1 or -1. By using
comparison methods, it is possible to prove that a small disturbance of the
stable dislocation evolves towards it. The same results hold for the stable
stationary solution $u_n(A,F)$ for $0<|F|<F_c$. As $|F|$ increases, a
disturbance of the stable stationary solution typically evolves towards the
same stationary solution displaced an integer number of lattice periods
unless the disturbance is sufficiently small. See Fig. \ref{fig12}(d) for an
example of this phenomenon for $F$ slightly smaller than $F_c$. Carefully
selecting the initial condition avoids this, as in Fig. \ref{fig12}(c).

\begin{figure}
\begin{center}
\includegraphics[width=10cm]{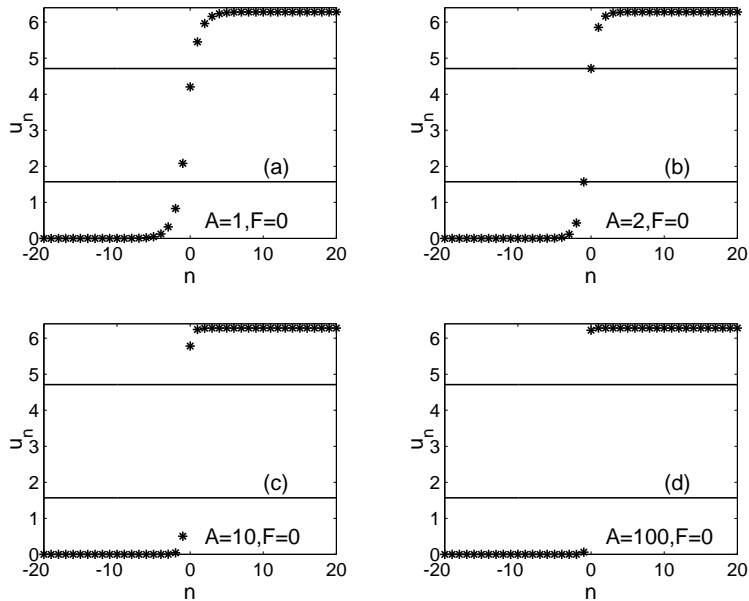}
\caption{Stationary solutions for the FK model when $A=1,2,10,100$.}
\label{fig3}
\end{center}
\end{figure}

For large $A$, the stable dislocation has $g'(u_n)>0$ for all $n$, and
(\ref{var}) gives $\lambda_1(A,0)>0$. Since $\lambda_1(A,F_c)=0$, this
implies that the critical field is nonzero. (Different proofs are given in
\cite{kee87,car00}). As $A>0$ decreases, several $u_n$ may enter the region
of negative slope $g'(u)$: the number of points with $g'(u_n)<0$ increases
as $A$ decreases, see Figures \ref{fig2} and \ref{fig3}. It should then be
possible to have $\lambda_1(A,0)=0$, i.e.\ $F_c=0$, for a discrete system!
Examples of this {\em pinning anomaly} will be given next.

\subsection{Pinning Failure}
Despite widespread belief, it is not true that the critical field is
positive for all discrete systems. This point was already raised by
Hobart \cite{hob65}, who proposed the following numerical
criterion to check whether for a given source $g$ the critical
field for (\ref{Fd}) is zero.

\begin{figure}
\begin{center}
\includegraphics[width=10cm]{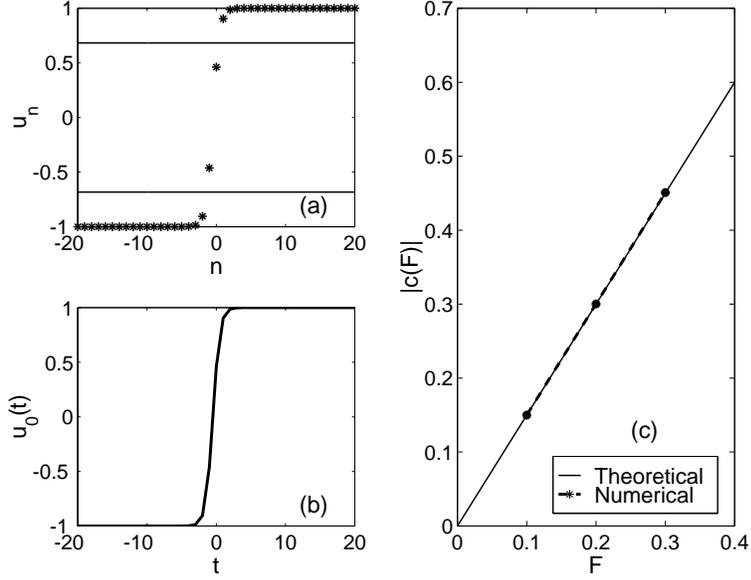}
\caption{(a) Stationary solution $u_n=\tanh(n)$. (b) wave front for $F$ small.
(c) Numerically calculated versus predicted speed for $g(u)=-2
\gamma u(1-u^2)/(1-\gamma u^2)$, with $\gamma=\tanh^2(1)$.}
\label{fig11}
\end{center}
\end{figure}

Let us assume for the sake of simplicity that $g$ is odd about $0$. Then
$U_2(0)=0$ and $U_1(0)= - U_3(0)$. For  any $x\in(U_1(0),U_3(0))$, we can
compute numerically a unique value $y(x)$ such that the sequence $u_n$
defined by $u_0 = x$, $u_1 = y(x)$ and $u_n = 2u_{n-1}-u_{n-2} +
g(u_{n-1}), n>1$ tends to $U_3(0)$ as $n \rightarrow \infty$. Hobart
conjectured that $F_c=0$ for a given nonlinearity $g$, provided that the
function $y(x)$ satisfies: 
\begin{eqnarray}
y^{-1}(x)=-y(-x), \quad y(x)-y(-x)=2x+g(x) , \label{hobart}
\end{eqnarray}
for $x\in (U_1(0),U_3(0))$. It is fairly easy to construct examples of
nonlinearities $g(x)$ for which (\ref{hobart}) holds. It suffices to
choose some smooth odd increasing function $u(x)$ such that $u(x)
\rightarrow \pm a$ as $x\rightarrow \pm \infty$ for some $a>0$. We
define $g(u(x))=u(x+1)-2u(x)+u(x-1)$ so that $g(z) = u(u^{-1}(z)+1)
-2z+u(u^{-1}(z)-1)$ and $y(z)=u(u^{-1}(z)+1)$. Choosing $u(x)= \tanh(x)$
\cite{sch79,bre97,fla99}, we get an explicit formula for $g$: $g(z)=-2 \gamma
z (1-z^2)/(1-\gamma z^2)$ with $\gamma=\tanh^2(1)$. Notice that one or two
points of the stationary solutions, $u_n=\tanh(n+p)$ ($p$ is any constant),
enter the region where $g'<0$; see Fig. \ref{fig11}(a).

By following this procedure, we find examples of bistable source terms
for which (\ref{Fd}) has a uniparametric family of continuous stationary
solutions, $u_n=u(n+p)$, $0\leq p<a$, satisfying $u_{n+1}-2u_n+u_{n-1} =
g(u_n)$ and $u_{-\infty}=-a$, $u_\infty = a$. In this case, (\ref{Fd})
does not have stationary solutions joining $U_1(F/A)$ and $U_3(F/A)$ unless
$F=0$ \cite{car99}. The existence of continuous steady solutions for $F=0$
implies that there is a continuous transition from wave fronts traveling to
the left ($c<0$) for $F>0$ to wave fronts traveling to the right ($c>0$) for
$F<0$. Only at $F=0$ wave fronts are stationary (pinned). This pinning
anomaly is stated more precisely as:

{\bf Theorem.} {\it Let $g\in C^2$ be as in the Introduction with $g(0)=0$,
and let ${\cal L}(F)$ be the operator
\begin{eqnarray}
{\cal L}(F)\, v_n = A\, g'(u_n) v_n + 2v_n - v_{n+1}-v_{n-1},
\label{LF}
\end{eqnarray}
corresponding to the evolution equation (\ref{Fd}) linearized about the
stationary solution $u_n=u_n(A,F)$ at field $F$. Let  us assume that 
for $F=0$, there exists a differentiable increasing stationary
solution $u(x)$, such that $u(x)\to\pm U_3(0)$ as $x\to\pm \infty$. Then,

1. Zero is the smallest eigenvalue of the operator ${\cal L}_0 = {\cal L}
(0)$, corresponding to the evolution equation (\ref{Fd}) linearized about
the stationary solution $u_n(A,0)=u(n)$.

2. $F_c(A)=0$ for (\ref{Fd}).

3. Traveling wave fronts exist for all $F\neq 0$. Furthermore, their speed
increases linearly with the force for small $F$. We have
\begin{equation}
c\sim -F\, {U_{3}(0)-U_{1}(0) \over \int_{-\infty}^{\infty}
\left({du\over dx}\right)^{2}\, dx},\label{cF}
\end{equation}
as $F\to 0$.

Moreover, Statement 3 implies the existence of steady differentiable
solutions $u(x)$ of (\ref{Fd}) such that $u(x)\to\pm U_3(0)$ as $x\to\pm
\infty$ for $F=0$.
 }

It is not our goal here to give a rigorous proof of this result, but to
sketch the main ideas. First of all, note that the derivative $v_n =
u_x(n)>0$ is a {\em positive} eigenfunction of the elliptic operator ${\cal
L}_0$ corresponding to the eigenvalue $\lambda=0$ and decaying exponentially
at infinity. Statement 1 immediately follows. This fact can be used to
construct propagating sub and supersolutions for (\ref{Fd}) which forbid
pinning for any $F\neq 0$. Thus,
$F_c=0$, which is Statement 2. For $F=\epsilon>0$ sufficiently small, the
propagating subsolutions are $l_n(t) = l(n + \epsilon c_0 t)$, with $c_0>0$
and $l(x) = u(x) + \epsilon u_x(x)$. For $F=-\epsilon$, the propagating
supersolutions are $w_n(t) = w(n-\epsilon c_0 t)$ with
$c_0>0$ and $w(x) = u(x) - \epsilon u_x(x)$. In both cases, we have to
choose $c_0 < 1/\mbox{max}\, (u_{x})$. A subsolution traveling to the
left ``pushes'' the fronts to the left. Similarly, the supersolutions
traveling to the right ``push'' the fronts to the right. 

Let us now obtain Statement 3. If $|F|>0$, we have traveling wave front
solutions $u_n(t)= {\cal U}(n-ct)$ of (\ref{Fd}), whose profile ${\cal
U}(z)$ satisfies the differential-difference equation:
\begin{equation}
- c\, {d{\cal U}\over dz}(z) = {\cal U}(z+1)-2{\cal U}(z)+ {\cal U}(z-1)
- g({\cal U}(z)) + F,
\end{equation}
and ${\cal U}(\pm\infty)=\pm U_3(0)$, \cite{car99}. Let $F= F_0\epsilon$
with $0<\epsilon\ll 1$. The traveling wave solution can be written as ${\cal
U}(n-ct) = u(n-ct) + \epsilon w(n-ct) + o(\epsilon)$, where $u(x)$ is the
smooth stationary profile. Let $z=n-c t$ and $c=c_0\epsilon + o(\epsilon)$.
Then $w$ obeys
\begin{eqnarray}
w(z+1)-2w(z)+w(z-1) -A g'(u(z)) w(z) =  -c_0 {du\over dz}(z) - F_0
,\nonumber \\ 
w(-\infty)=w(\infty)= {1\over g'(U_1)}={1\over g'(U_3)}\,. \nonumber
\end{eqnarray}
By the Fredholm alternative, this linear nonhomogeneous equation has a
solution if the left hand side $-c_0 du/dz- F_0$ is orthogonal to the
eigenfunction $du/dz$, which yields (\ref{cF}).

In Section \ref{sec:asymptotic}, we show that the wave front speed $c$
scales as $|F-F_c|^{{1\over 2}}$ if $F_c>0$. Our linear scaling
(\ref{cF}) of the velocity in Statement 3 therefore implies that $F_c=0$.
The linear scaling (\ref{cF}) with $F_c=0$ implies
the existence of smooth stationary solutions at $F=0$ as discussed in
the first Subsection. 
\bigskip

\noindent 
{\it Remark 1.} We conjecture the three statements in the above theorem
are equivalent. To prove this, it would be enough to show that $F_c=0$
implies the linear scaling of the speed of the waves (Statement 3). Then,
existence of differentiable stationary solutions follows. This implies
statement 1 ($\lambda_1(A,0)=0$), which implies Statement 2 ($F_c=0$), 
as we showed above.  
\bigskip

\noindent 
{\it Remark 2.}  When stationary wave front solutions have smooth profiles,
pinning failure occurs for discrete RD equations and for discrete equations
with conservative dynamics. In the latter case, translation invariant smooth
profiles have the same energy and therefore the PN energy barrier (defined
as the smallest energy barrier to overcome for a kink or wave front to move
\cite{bra98}) vanishes. Pinning of a wave front usually results if the
energy difference between the stable and the unstable front solutions (see
Fig.~\ref{fig2}) is not zero. This energy difference provides an estimation
of the PN energy barrier. Discusions of the PN potential and the PN barrier
can be found in Section 2.3 of Ref.Ê\cite{bra98} and in Section III.B of Ref.
\cite{fla99}. The mathematical meaning and usefulness of the PN barrier for
an infinite system with conservative dynamics are worth studying. 
\bigskip 

\noindent 
{\it Remark 3.} Speight and Ward \cite{spe94} and Speight
\cite{spe97,spe99} have developed a technique to discretize some continuum
conservative models in such a way that kink-like initial profiles may
propagate without getting trapped. Their idea is to seek a discrete version
of the potential energy which admits minimals satisfying a first order
difference equation called the Bogomol'nyi equation so that there is no PN
barrier. On the other hand, the difference operators in Speight et al
discretized equations of motion have a structure different from discrete
diffusion and are hard to justify physically.
\bigskip 

\noindent 
{\it Remark 4.} In discrete RD equations, moving and pinned fronts cannot
coexist for the same value of the applied field. For chains with
conservative Hamiltonian dynamics, the situation is less clear. In fact, it
is possible to have two stationary front solutions with a positive energy
difference between them (which would imply a nonzero critical field and
therefore wave front pinning according to general belief), and yet a moving
wave front may coexist with the stationary fronts for the same parameter
values. An explicit example of this situation has been constructed by Flach,
Zolotaryuk and Kladko using an inverse method \cite{fla99}. %Thus naive
%energy arguments alone are not enough to rule out the existence of moving
%wave fronts for chains with conservative Hamiltonian dynamics. 

\setcounter{equation}{0}
\section{Asymptotic theory of wave front depinning}
\label{sec:asymptotic} 
In this section we introduce a systematic procedure to derive analytic
expressions for the critical field $F_c>0$ as a function of $A$, and for the
front profiles and their velocity as functions of $F-F_c$ and $A$. Our
methods work best in the strongly discrete case, for large $A$. Our
ideas are quite general and may be applied successfully to more complex
discrete models \cite{abc}. We shall assume that $g\in C^2$ throughout this
Section. 

\subsection{Theory with a single active point}
\begin{figure}
\begin{center}
\includegraphics[width=6cm]{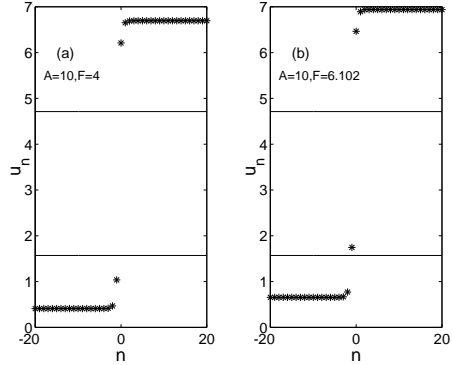}
\caption{Stationary solutions for the FK model with $A=10$:
(a) No points are found in the region $g'<0$ for sufficiently small $F$; (b)
one point enters the region $g'<0$ for sufficiently large $F<F_c$.}
\label{fig4}
\end{center}
\end{figure}

We choose $A$ large enough for the stable dislocation in Fig. \ref{fig2}
not to enter the region where $g'<0$, see Fig. \ref{fig3}. When $F>0$,
this solution is no longer symmetric with respect to $U_2$. If $F$ is
not too large, all $u_n(A,F)$ avoid the region of negative slope $g'(u)
<0$. For larger $F$ and generic potentials (FK, double-well,\ldots), we have
observed numerically that $g'<0$ for a single point, labelled $u_0(A,F)$.
This property persists until $F_c$ is reached, see Fig. \ref{fig4}.

\begin{figure}
\begin{center}
\includegraphics[width=10cm]{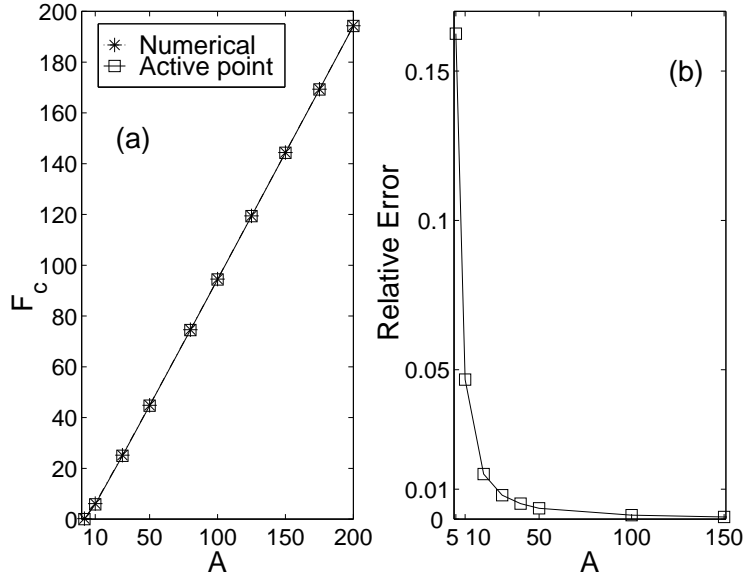}
\caption{Approximation of Eq.(\ref{Fd}) by the equation with one active point
for the FK potential and $A>2$: (a) Critical force versus $A$.
(b) Error in the approximation of $F_c(A)$. }
\label{fig5}
\end{center}
\end{figure}

Firstly, consider the symmetric stationary profile with $u_n\neq U_2$ for
$F=0$. The front profile consists of two tails with points very close to
$U_1$ and $U_3$, plus two symmetric points $u_0$, $u_1$ in the gap region
between $U_1$ and $U_3$. As $F>0$ increases, this profile changes slightly:
the two tails are still very close to $U_1(F/A)$ and $U_3(F/A)$. As for the
two middle points, $u_1$ gets closer and closer to $U_3$ whereas $u_0$ moves
away from $U_1$. This structure is preserved by the traveling fronts above
the critical field: there is only one active point most of the time, which
we can adopt as our $u_0$. Then the wave front profile (\ref{e2}) can be
calculated as $u(-ct)= u_0(t)$. In Eq.\ (\ref{Fd}), we can approximate
$u_{-1}\sim U_1$, $u_1\sim U_3$, thereby obtaining
\begin{eqnarray}
{du_{0}\over dt}\approx U_1\left({F\over A}
\right) + U_3\left({F\over A} \right)-2 u_0 -
A\, g(u_0) + F .\label{u0}
\end{eqnarray}
This equation has three stationary solutions for $F<F_c$, two stable and one
unstable, and only one stable stationary solution for $F>F_c$. Let us
consider $F<F_c$. Only two out of the three solutions of Eq.\ (\ref{u0})
approximate stationary fronts for the exact system: those having smaller
values of $u_0$. The one having smallest $u_0$ approximates the stable
stationary front, the other one approximates the unstable stationary front.
Recall that the unstable front had a value $u_0 = [U_1(0) + U_3(0)]/2$ at
the middle of the gap for $F=0$. As $F>0$ increases, $u_0$ decreases towards
$U_1(F/A)$. Thus one active point will also approximate the profile of the
unstable stationary front. The stationary solution of Eq.\ (\ref{u0}) having
the largest value of $u_0$ (slightly below $U_3(F/A)$) is not consistent with
the assumptions we made to derive Eq.\ (\ref{u0}), and therefore it does not
approximate a physically existing stationary front. If $F>F_c$, the only
stationary solution of Eq.\ (\ref{u0}) is the unphysical one. The critical
field $F_c$ is such that the expansion of the right hand side of (\ref{u0})
about the two coalescing stationary solutions has zero linear term, $2 +A
g'(u_0)=0$, and
\begin{eqnarray}
2 u_0 + A \, g(u_0) \sim  U_1\left({F_c\over A}
\right) + U_3\left({F_c\over A} \right) + F_c .
\label{Fc1}
\end{eqnarray}
These equations for $F_c$ and $u_0(A,F_c)$ have
been solved for the FK potential, for which $u_0 =
\cos^{-1}(-2/A)$ and $U_1+U_3 = 2\sin^{-1}
(F_c/A) + 2\pi$. The results are depicted in Fig.
\ref{fig5}, and show excellent agreement with
those of direct numerical simulations for $A>10$.
Our approximation performs less well for smaller
$A$, and it breaks down at $A=2$ with the wrong
prediction $F_c=0$. Notice that $F_c(A)/A\sim 1$ as
$A$ increases. In practice, only steady solutions
are observed for very large $A$.

\begin{figure}
\begin{center}
\includegraphics[width=10cm]{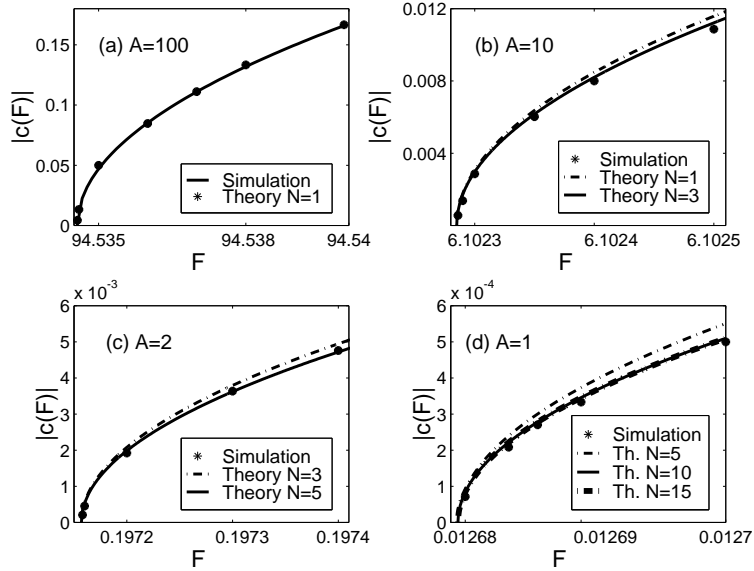}
\caption{Comparison of theoretically predicted and numerically
calculated wave front velocities near $F_c$ for the FK model with $N$
active points and the following values of the parameter $A$: (a) $A=100$,
(b) $A=10$, (c) $A=2$, (d) $A=1$.}
\label{fig6}
\end{center}
\end{figure}

Let us now construct the profile of the traveling
wave fronts after depinning, for $F$ sligthly above
$F_c$. Then $u_0(t)= u_0(A,F_c) + v_0(t)$ obeys
the following equation:
\begin{eqnarray}
{dv_{0}\over dt} &=& \alpha\, (F-F_c) + \beta\, v_{0}^{2},
\label{v0}\\
\alpha &=& 1 + {1\over A\, g'(U_{1}(F_{c}/A))} + {1\over A\,
g'(U_{3}(F_{c}/A))}\,,\label{alpha0}\\
\beta &=& -{A\over 2}\, g''(u_0),\label{beta0}
\end{eqnarray}
where we have used $2+Ag'(u_0)=0$, (\ref{Fc1}) and ignored terms of order
$(F-F_c)\, v_0$ and higher. These terms are negligible after rescaling
$v_0= (F-F_c)^{{1\over 2}}\varphi$ and $\tau= (F-F_c)^{{1\over 2}}t$. The
coefficients $\alpha$ and $\beta$ are positive because $g'(U_i)>0$ for $i=1,
3$ and $g''(u_0)<0$ since $u_0\in (U_1(0),U_2(0))$. For the FK potential,
$\alpha=1+ 2/\sqrt{A^2 -F_c^{2}}$ and $\beta= \sqrt{A^2 -F_c^{2}}/2$.
Equation (\ref{v0}) has the (outer) solution
\begin{eqnarray}
v_0(t)\sim \sqrt{{\alpha (F-F_{c})\over \beta}}\,
\tan\left(\sqrt{\alpha\beta\, (F-F_{c})}\, (t-t_0)\right)\,,
\label{outer}
\end{eqnarray}
which is very small most of the time, but it
blows up when the argument of the tangent function
approaches $\pm \pi/2$. Thus the outer
approximation holds over a time interval
$(t-t_0)\sim \pi/\sqrt{\alpha\beta\, (F-F_{c})}$, which equals
$\pi\sqrt{2/\alpha} (A^2-4)^{ -{1\over 4}} (F-F_c)^{-{1\over 2}}$ for
the FK potential. The reciprocal of this time interval
yields an approximation for the wave front velocity,
\begin{eqnarray}
c(A,F)\sim - {\sqrt{\alpha\beta\, (F-F_{c})}\over \pi}\,,
\label{c}
\end{eqnarray}
or $c\sim -(A^2-4)^{{1\over 4}} (1+2/\sqrt{A^2 -F_c^{2}})^{{1\over
2}} (F-F_c)^{{1\over 2}} /(\pi\sqrt{2})$ for a FK potential. The minus sign
reminds us that wave fronts move towards the left for $F>F_c$. In Figs.\
\ref{fig6}(a) and (b) we compare this approximation with the numerically
computed velocity for $A=100$ and $A=10$. 

\begin{figure}
\begin{center}
\includegraphics[width=10cm]{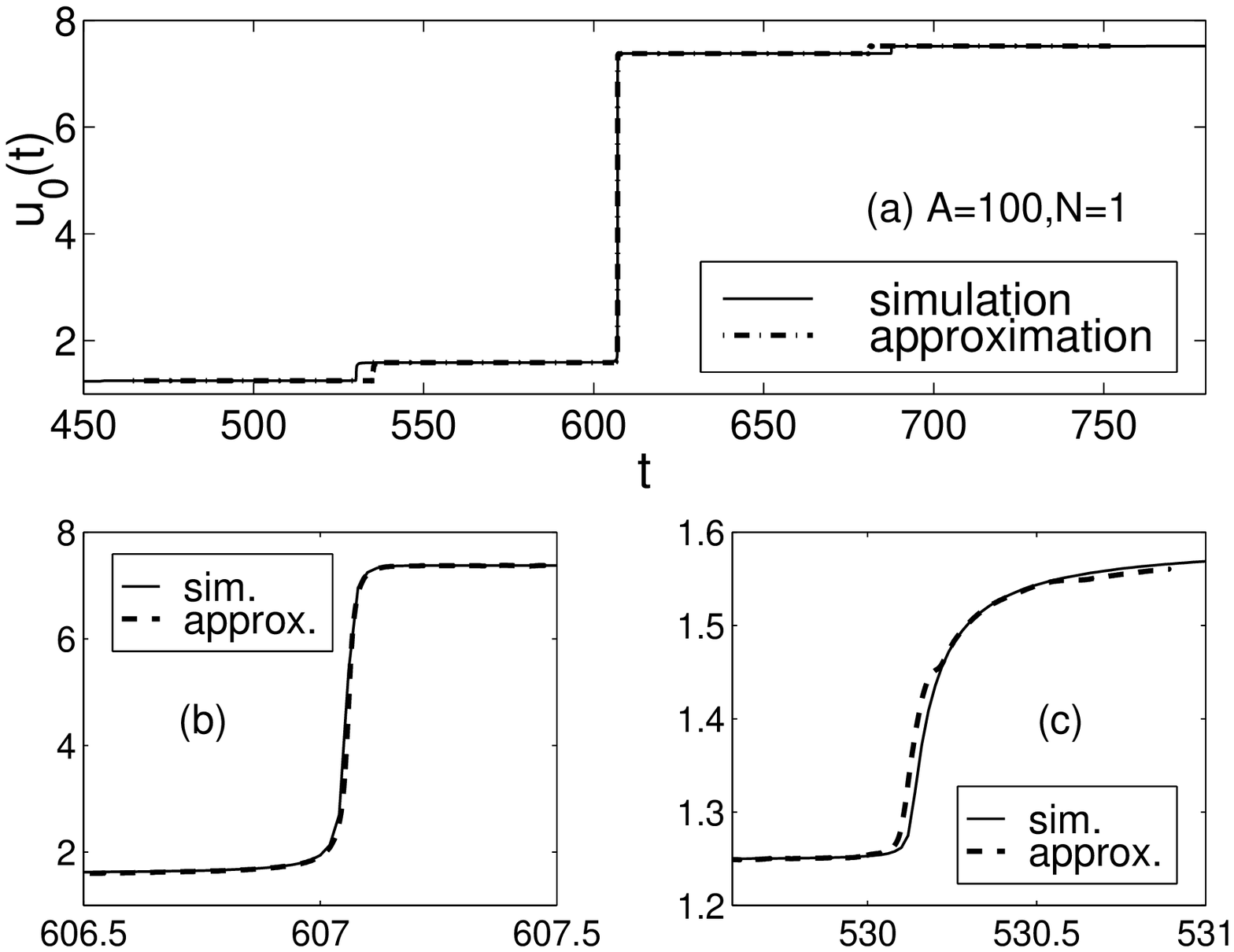}
\caption{Comparison of asymptotic and numerically
calculated wave front profiles near $F_c$:
(a) Complete wave front profile as indicated by the trajectory $u_0(t)$.
(b) Zoom near the largest jump in the profile.
(c) Zoom near the jump preceding the largest one after translating the
asymptotic profile. The latter has been calculated by inserting the
approximate $u_0(t)$ in the equation for $u_{-1}(t)$.}
\label{fig7}
\end{center}
\end{figure}

When the solution begins to blow up, the outer solution (\ref{outer}) is no
longer a good approximation, for $u_0(t)$ departs from the stationary value
$u_0(A,F_c)$. We must go back to (\ref{u0}) and obtain an inner approximation
to this equation. As $F$ is close to $F_c$ and $u_0(t)-u_0(A,F_c)$ is of
order 1, we solve numerically (\ref{u0}) at $F=F_c$ with the matching
condition that $u_0(t)-u_0(A,F_c)\sim 2/[\pi\sqrt{\beta/[\alpha\, (F-F_c)]}
- 2\beta\, (t-t_0)]$, as $(t-t_0)\to -\infty$. This inner solution describes
the jump of $u_0$ from $u_0(A,F_c)$ to values on the largest stationary
solution of Eq.\ (\ref{u0}), which is close to $U_3$. During this jump, the
motion of $u_0$ forces the other points to move. Thus, $u_{-1}(t)$ can be
calculated by using the inner solution in (\ref{Fd}) for $u_0$, with $F= F_c$
and $u_{-2}\approx U_1$. A composite expansion \cite{bon87} constructed with
these inner and outer solutions is compared to the result of direct
numerical simulations in Fig. \ref{fig7}. 

Notice that (\ref{v0}) is the normal form
associated with a saddle-node bifurcation in a
one dimensional phase space. The wave front
depinning transition is a {\em global} bifurcation
with generic features: each individual point
$u_n(t)$ spends a long time, which scales as
$|F-F_c|^{-{1\over 2}}$, near discrete values
$u_n(A,F_c)$, and then jumps to the next discrete
value on a time scale of order 1. The traveling
wave ceases to exist for
$F\leq F_c$. 

\subsection{Theory with several active points}
The approximations to $F_c(A)$ and the wave front
speed provided by the previous asymptotic theory
break down for small $A$. In
particular, for the FK potential and $A<2$, no
double zeroes of $2x +A\sin(x) - (F +U_{1} +U_3)$
are found for $F=F_c$. What happens is that we
need more than one point to approximate wave front
motion. Depinning is then described by a reduced
system of more than one degree of freedom
corresponding to active points. There is a
saddle-node bifurcation in this reduced system
whose normal form is of the same type as
(\ref{v0}). The jump of the active points after
blow up is found by solving the reduced system
with a matching condition \cite{CB}.

We explain our procedure for a finite number of active points.
The front is formed by two tails, very close to
$U_1(F/A)$ and $U_3(F/A)$ respectively, and several intermediate
points $u_{-L},\ldots,u_M$. The reduced system describing the dynamics
of the front for $F>F_c$ is:
\begin{eqnarray}
{du_{i}\over dt} = u_{i+1}-2u_i+u_{i-1}-Ag(u_i)+F, \quad i=-L,\ldots,M,
\nonumber \\
u_{-L-1}=U_1(F/A), \; u_{M+1}=U_3(F/A) \label{red1}
\end{eqnarray}
At $F=F_c$ this system has a stationary solution $u_{-L},...u_M$.
Writing $u_i(t)=u_i+ v_i(t)$, we obtain for $v_i$,
\begin{eqnarray}
\quad\quad {dv_{i}\over dt} =v_{i+1}-2v_i+v_{i-1}-Ag'(u_i) v_i - {A\over 2}
g''(u_i) v_i^2 + F - F_c, \; i=-L,\ldots,M \label{red2} \\
v_{-L-1}\sim {F-F_c \over A g'(U_1(F_c/A)) }\,, \quad
v_{M+1}\sim {F-F_c \over A g'(U_3(F_c/A)) }\,. \nonumber
\end{eqnarray}
The tridiagonal matrix ${\cal M}$ defined as
\begin{eqnarray}
\left( \begin{array}{ccccc}
\!\!\!\! 2+Ag'(u_{-L}) & -1  &    &   &  \\
-1    & \!\!\!\! 2+Ag'(u_{1-L})& -1\hskip .8cm  & &  \\
 &  .\hfill &  .\hfill &  .\hfill &        \\
& . & . & . & \\
& \hfill . & \hfill . & \hfill . & \\
 &  & \hskip .8cm -1  & 2+Ag'(u_{M-1})\!\!\!\! & -1 \\
& &   & -1 &  2+Ag'(u_{M}) \!\!\!\!\!
\end{array} \right)\label{red3}
 \end{eqnarray}
has $L+M$ strictly positive eigenvalues plus a smallest eigenvalue
$\lambda \sim 0$ with an associate positive eigenfunction $V$.
We choose $V$ such that $\sum_{i=-L}^M V_i^2=1$
and write an eigenfunction expansion for $v$:
\begin{eqnarray}
v(t) = \varphi\, V +\sum_{i=-L, i\neq 0}^M W_i \varphi_i \exp (-\lambda_i
t) .  \label{red4}
\end{eqnarray}
Thus $v(t) \sim V \varphi(t)$ as time increases. Let
$D=-{A\over 2}$ diag $( g''(u_{-L}),\ldots, g''(u_M))$,
$V^2=(V_{-L}^2, \ldots, V_M^2)$ and
$$w= (F-F_c) \left(1+ {1 \over A g'(U_1(F_c/A)) },1,...,1,1+ {1 \over
A g'(U_3(F_c/A)) }\right).$$
Then system (\ref{red2}) becomes
\begin{eqnarray}
{d\varphi\over dt}\, V \sim  \varphi\, {\cal M}V  + \varphi^2\, D V^2 +
w \sim \varphi^2\, D V^2 + w,
\label{red5}
\end{eqnarray}
as time increases. Multiplying by the transpose of $V$, we get
an evolution equation for the amplitude $\varphi(t)$:
\begin{eqnarray}
{d\varphi\over dt} = \alpha\, (F-F_c) +\beta\,\varphi^2,
\label{red6}
\end{eqnarray}
where now $\alpha$ and $\beta$ are:
\begin{eqnarray}
\alpha= \sum_{i=-L}^M V_i +{V_{-L} \over A g'(U_1(F_c/A)) }
+ {V_M \over A g'(U_3(F_c/A)) }>0,\nonumber\\
\beta=-{A\over 2}\sum_{i=-L}^M g''(u_i)V_i^3 >0.\nonumber
\end{eqnarray}
The coefficient $\alpha$ is positive because $g'(U_i)>0$ for $i=1,3$. We
have checked numerically that $\beta>0$ for different nonlinearities
and values of $A$. An intuitive explanation follows. First of all, notice
that $g''(u)>0$ for $u\in (U_2(0),U_3(0))$ and $g''(u)<0$ for $u\in
(U_1(0),U_2(0))$. For large $A$, the largest component is $V_0$, the others
are negligible and we have one active point as in the previous subsection;
see Fig.\ \ref{fig7}. Then $\beta\sim -g''(u_0(A,F_c)) V_0^3>0$ because
$u_0<U_2(0)$, which implies $g''(u_0)<0$. As $A$ decreases, $V_0$ is still
the largest component and $g''(u_0(A,F_c))<0$. Now there may be other terms
with $g''(u_i(A,F_c))>0$ and we only have numerical evidence that $\beta>0$,
but not a proof. 

Notice that (\ref{red6}) is the normal form of a saddle-node bifurcation. Its
solution is again (\ref{outer}), which blows up at times $(t-t_0) =\pm
1/(2c)$, where 
\begin{equation}
c(A,F)\sim -{1\over \pi}\sqrt{\alpha \beta (F-F_c)}, \label{red7}
\end{equation}
as discussed before. $c$ is the wave front speed near $F_c$, approximately
given by the reciprocal of the time during which the outer solution holds.

\begin{figure}
\begin{center}
\includegraphics[width=10cm]{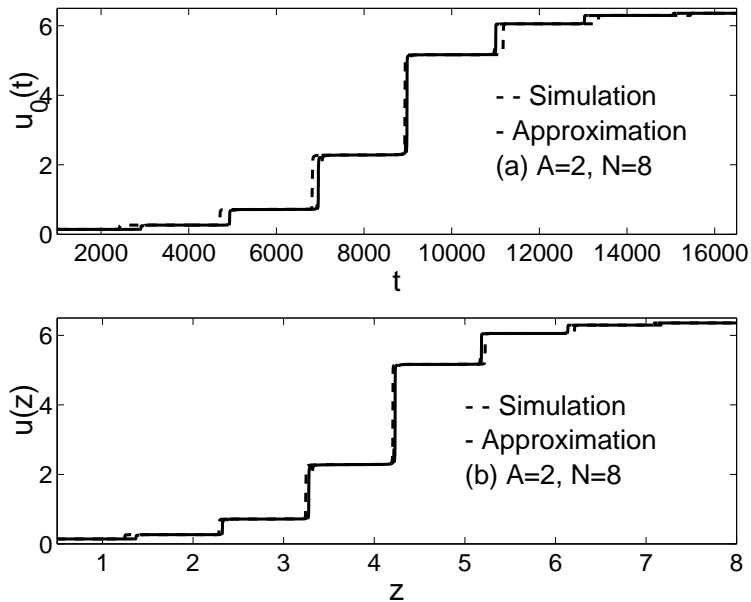}
\caption{Comparison of theoretically predicted and numerically 
calculated wave fronts near $F_c$ for $A=2$ using $N=8$ active points:
(a) Trajectory of one point. (b) wave front profile, $u(z)=u_0(z/|c|)$.}
\label{fig9}
\end{center}
\end{figure}

Figures \ref{fig10}, \ref{fig6} and \ref{fig9} show the critical field,
wave front velocities and profiles for different values of $A\in (1,10)$
corresponding to the FK model. We have compared results of direct
numerical simulations to those of our theory for $N=L+M+1$ active
points. Provided $N=L+M+1$ active points have been selected, we find the
smallest eigenvalue of the matrix ${\cal M}$ and move $F$ until $\lambda(F,A
;N) =0$, $N=L+M+1$, which yields an approximation for $F_c(A)$. See Fig.\
\ref{fig10}. The wave front velocities can be calculated by means of Eq.\
(\ref{red7}) and have been depicted in Fig. \ref{fig6}. 

The wave front profiles near $F_c$ can be determined as follows. We start
with an initial condition, $u_n(0)\approx u_n(A,F_c)$, or $\varphi(0)=0$ in
Eq.\ (\ref{red4}). The active points blow up at $t\sim \pm (2c)^{-1}$, for
example as 
\begin{equation}
u_n(t) \sim u_n(A,F_c) + {1\over \beta(\pm {1\over 2c }- t)}\,
V,\label{red9} 
\end{equation}
provided $t\to\pm 1/(2c)$. At these times, we should insert a fast stage
during which the $u_n(t)$ are no longer close to $u_n(A,F_c)$, as an inner
layer. The inner layer variables $u_n(t)$ obey Eq.\ (\ref{Fd}) with $F=F_c$
and the boundary conditions $u_n(t)\to u_n(A,F_c)$ [according to Eq.\
(\ref{red9})] as $t\to - \infty$ and $u_n(t)\to u_{n+1}(A,F_c)$ as $t\to
\infty$. To get a uniform approximation, we notice that the blow up times
are $t_m = (2c)^{-1} +m/c$, $m\in \ZZ$. Let us denote by $u_n^{(m)}(\tau)$,
$\tau= (t-t_m)$, the solution of Eq.\ (\ref{Fd}) with $F=F_c$ and the
boundary conditions $u_n^{(m)}(\tau)\to u_{n+m}(A,F_c)$ as $\tau\to - \infty$
and $u_n^{(m)}(\tau)\to u_{n+m+1}(A, F_c)$ as $\tau\to\infty$. During the 
time interval $(t_{-L-n-1},t_{M-n}) =(-(2c)^{-1}-(L+n)/c, (2c)^{-1}+
(M-n)/c)$, that $u_n(t)$ needs to go from $U_1(F_c/A)$ to $U_3(F_c/A)$, the
uniform approximation to the wave front is 
\begin{eqnarray}
u_n(t) \sim \sum_{m=-n-L-1}^{M-n} \left\{ u_n^{(m)}(t-t_m)
+u_n^{(m-1)}(t-t_{m-1}) - u_{n+m}(A,F_c)\right.\label{red10}\\
\left. + \left[\varphi \left(t-{m\over c}\right) - {1\over \beta\,
(t_m-t)}+ {1\over \beta\, (t-t_{m-1})}\right]\, V\right\}
\chi_{(t_{m-1},t_{m})}. \nonumber 
\end{eqnarray}
Then $u_n(t_{-L-n-1})\sim U_1(F_c/A)$ and $u_n(t_{M-n}) \sim U_3(F_c/A)$. In
Eq.\ (\ref{red10}), the indicator function $\chi_{(t_{m-1},t_{m})}$ is 1 if
$t_{m-1}<t<t_{m}$ and 0 otherwise. Therefore, $\chi_{(t_{m-1},t_{m})}=
\theta(t_m-t)- \theta(t_{m-1}-t)$, where $\theta(x)=1$ if $x>0$ and 0
otherwise. Written in terms of the variable $z=n- ct$, such that $u_n(t)=
u(z)$, $u(z) = u_n((n-z)/c) = u_0(-z/c)$. Then Eq.\ (\ref{red10}) becomes
\begin{eqnarray}
\quad\quad\quad u(z) \sim \sum_{m=-L-1}^{M} \left\{
u_0^{(m)}\left(-{z+m+{1\over 2}\over c}\right) +
u_0^{(m-1)}\left(-{z+m-{1\over 2}\over c}\right) -u_m(A,F_c) \right.
\label{red11}\\
\left. + \left[\varphi \left(-{z+m\over c}\right) - {c\over \beta\,
(z+m+{1\over 2})}- {c\over \beta\, (z+m-{1\over 2})}\right]\,
V\right\}\nonumber\\
\times \left[\theta\left(z+m+{1\over 2}\right)-\theta\left(z+m - {1\over 2}
\right)\right],  \nonumber
\end{eqnarray}
for $-M-1/2<z< L + 1/2$. We have $u(L+1/2+0)\sim U_1(F_c/A)$ and
$u(-M-1/2-0)\sim U_3(F_c/A)$, and therefore Eq.\ (\ref{red11}) approximates
the wave front profile. In Fig.\ \ref{fig9}, we have depicted the wave front
profile in two ways, by drawing $u_0(t)$ and $u(z)= u_0(-z/c)$. Notice that
the largest source of discrepancy between numerical calculations and our
asymptotic approximation is the error in determining the wave speed. The
discrepancies are more evident for $u_0(t)$ because of the different
horizontal scale used to depict $u(z)$. 

How do we determine the optimal number of active points? For large
enough $N=L+M+1$ and a given $A$, the eigenvector $V$ corresponding to the
smallest eigenvalue of the matrix ${\cal M}$ in (\ref{red3}) has a certain
number of components that are of order one, whereas all others are very
small. The number of components of normal size determines the optimal
number of active points: only one point if $A$ is larger than 10, five
if $A=2$, etc. Keeping less active points than the optimal number results in
larger errors whereas keeping more active points than optimal does not
result in a significantly better approximation. The eigenvector of the
reduced system of equations for the active points is a good approximation to
the large components of the eigenvector corresponding to the complete
system. As we approach the continuum limit, more and more points enter the
reduced system of equations and exponential asymptotic methods become a
viable alternative to our methods.

\subsection{Depinning transition as a global bifurcation}
We have shown that the depinning transition is a global bifurcation in a
reduced system of equations corresponding to the active points. Starting
from a stable stationary solution, the smallest eigenvalue of the system 
linearized about the stationary solution becomes zero at the (approximate)
critical field, and its associated eigenfunction is positive. The stationary
solution disappears as the critical field is surpassed. Beyond it, the
active points $u_n(t)$ spend a long time, of order $(F-F_c)^{-{1\over 2}}$,
near the stationary values $u_n(A,F_c)$, and then jump to $u_{n+1}(A,F_c)$
on an order 1 time scale. Near the critical field, the depinning transition
is described {\em locally} by the normal form of a saddle-node bifurcation.
For $F>F_c$ (or $F<-F_c$), the bifurcation amplitude blows up in finite
time, on a time scale of order $|\, |F|-F_c |^{-{1\over 2}}$. The
construction of the wave front profile is completed matching the outer
solution given by the saddle-node normal form to a solution of the reduced
system of active points at $F=F_c$ (a mathematically related phenomenon
occurs in a mean-field model of sliding charge-density waves \cite{bon87}). 

We conjecture that the depinning transition in the infinite system (\ref{Fd})
is a global bifurcation of the same type as for the reduced system of active
points. At the critical field, two stationary solutions of Eq.(\ref{Fd})
(one stable, the other unstable) coalesce and disappear. For $F>F_c$ (or
$F<-F_c$), the wave front profile is constructed as indicated above for the
reduced system. To prove this conjecture, we could repeat our construction
in Section 3.2 for an infinite number of points. This is possible because we
know that the infinite system, linearized about the `stable' steady solution
$u_i(A,F_c)$ at $F=F_c$, has a zero eigenvalue and an associated positive
exponentially decaying eigenfunction $V$. Using $V$, we obtain the normal
form equation (\ref{red6}), where now 
\begin{eqnarray}
\alpha=\sum_{i=-\infty}^{\infty} V_i >0, \quad \beta=
-{A\over 2}\sum_{i= -\infty}^{\infty} g''(u_i)V_i^3 .\label{red8}
\end{eqnarray} 
We should now prove that the coefficient $\beta$ is positive and that the
infinite system has solutions connecting $u_n(A,F_c)$ to $u_{n+1}(A,F_c)$
and satisfying the matching condition. We justified that $\beta>0$ for the
finite system in Subsection \ref{sec:asymptotic}.2 and we show in
Proposition A.1 (Appendix A) that the eigenfunction for the infinite system
can be approximated by the corresponding eigenfunction of the reduced system
with a finite number of active points. The existence of traveling wave
solutions for $F>F_c$ ensures that the infinite system has solutions
connecting $u_n(A,F_c)$ to $u_{n+1}(A,F_c)$. The velocity of a wave front in
the infinite system is again given by (\ref{red7}) with the coefficients 
(\ref{red8}).
\bigskip

\noindent {\it Remark 5.} By using comparison techniques, it is possible to
prove that solutions of discrete RD equations with finitely many points and
Dirichlet boundary conditions approximate solutions of the same equations
with infinitely many points. In the continuum limit, the wave fronts approach
constant values exponentially fast as $i\to\pm\infty$. This exponential
decay justifies the active point approximation in two ways. First of all,
the number of active points needed to approximate well the wave fronts of
the infinite system decreases as $A$ increases. It is usually better to
add another active point to the approximate system than to patch rigid tails
to the last active points of a wave front by generalizing Kladko et al's
active site approximation \cite{kla00}. Secondly, exponential decay at the
ends of a wave front causes the operator of the linearized problem about the
wave front to be compact, and therefore to have discrete spectrum (see
Appendix A). This fact justifies that the normal form we calculate by using
active points approximates the correct local normal form of the depinning
global bifurcation.

\setcounter{equation}{0}
\section{Conclusions}
In this paper, we have studied depinning of wave fronts in discrete RD
equations. Pinned (stationary) and traveling wave fronts cannot coexist
for the same value of the forcing term. There are two different
depinning transitions, i.e.\ two different ways in which a pinned
front may start moving. The normal depinning transition can be viewed
as a loss of continuity of traveling front profiles as the critical
field is approached: below the critical field, the fronts become
pinned stationary profiles with discontinuous jumps at discrete values
$u_n$. The wave front velocity scales as $|F-F_c|^{{1\over 2}}$ near
the critical field $F_c$. For sufficiently large $A$ (far from the
continuum limit), the critical field and these fronts can be
approximated by singular perturbation methods which show excellent
agreement with numerical simulations. These methods are based upon
the fact that the wave front motion can be described by a reduced
system of equations corresponding to the dynamics of a finite number
of points only: the active points.

Besides the normal depinning transition, certain nonlinearities
present anomalous pinning (pinning failure): the velocity of the
wave fronts is not zero except at zero forcing, just as for continuous
RD equations. These nonlinearities are characterized by smooth
profiles of stationary and moving wave fronts, by having zero critical
field and by a linear scaling of wave front velocity with field.

\renewcommand{\theequation}{A.\arabic{equation}}
\setcounter{equation}{0}
\appendix
\section{Characterization of the depinning threshold}

In this section we establish the `depinning criterion' which
provides a characterization of $F_c(A)$:

{\bf Theorem A.1}
{\it  Set $F=0$ and $A>0$. Assume that the nonlinearity $g \in C^3$ has
three zeroes $U_i$, $U_1<U_2<U_3$, is odd about $U_2$ and satisfies
$g'(U_1)=g'(U_3)>0$.  Let $u_n$ be a stationary increasing solution of
(\ref{Fd}), symmetric about $U_2$ and such that $u_{-\infty}=U_1$ and
$u_{\infty}=U_3$. Let $\lambda_1(A,0)$ be the smallest eigenvalue of the
zero field operator ${\cal L}_0$ of (\ref{LF}) at $F=0$:
\begin{eqnarray}
-(v_{n+1}-2v_n+v_{n-1})+ Ag'(u_n) v_n = \lambda_1(A,0) v_n
\nonumber\\
v_{\pm n} \rightarrow 0 \hbox{ exponentially as $n\to \infty$}
\label{lin0}
\end{eqnarray}
If $\lambda_1(A,0)>0$, then $F_{c}(A)>0$ and for $|F|\leq F_{c}(A)$ there
exist increasing stationary solutions $u_n(A,F)$ of (\ref{Fd}) with
$u_{-\infty}=U_1(F/A)$ and $u_{\infty}=U_3(F/A)$. Moreover, the smallest
eigenvalues of the operator ${\cal L}(F)$, corresponding to the
linearization of (\ref{Fd}) about $u_n(A,F)$,
\begin{eqnarray}
-(v_{n+1}-2v_n+v_{n-1})+ Ag'(u_n(A,F)) v_n= \lambda_1(A,F) v_n
\nonumber\\
v_{\pm n} \rightarrow 0 \hbox{ exponentially as $n\to \infty$,}
\label{linF}
\end{eqnarray}
are strictly positive for $|F|<F_{c}(A)$. We can characterize $F_{c}(A)$ as
the zero of the smallest eigenvalue, $\lambda_1(A,F_{c}(A))=0$.
}

This theorem will be proven in subsection A.2. To calculate
$\lambda_1(A,F)$ and $F_{c}(A)$, we approximate the infinite tridiagonal
matrix in Eq.\ (\ref{linF}) by a $N\times N$ matrix, where $N$ is the number
of active points. Similar truncation approximations were used in
\cite{mil99} to calculate the lowest eigenvalue of an infinite tridiagonal
matrix. For values of $A$ which are not too small, numerical simulations
show that the matrices ${\cal M}$ in (\ref{red3}) have positive eigenvalues.
The eigenvector $V(A,F,N)$, (chosen to have norm $1$) corresponding to the
smallest eigenvalue, $\lambda(A,F,N)$, is positive and it is `concentrated'
in the central components, $V_{-m(A)},\ldots,V_{m(A)}$. All other components
are very small. The number of significant components $m(A)$ does not change
as $N$ increases, but it increases as $A$ decreases. For large $A$, $m(A)=0$
and only $V_0$ is significant. Provided $N$ is large enough, the eigenvalues
$\lambda(A,F,N)$ and the eigenvectors $V(A,F,N)$ approximate well the
smallest eigenvalue and associated eigenfunction of the infinite problem, as
we indicate in the next subsection. For a fixed value of $N$, the eigenvalues
$\lambda(A,F,N)$ decrease as $A$ decreases. For fixed $N$ and $A$, they
decrease as $F$ increases from $F=0$ to values close to $F_c(A)$. In the next
subsection, we collect several results on eigenvalues for this type of
problems.

\subsection{Eigenvalue problems}
Before proving Theorem A.1, we should make sure that our linear operators do
have  eigenfunctions and eigenvalues. We consider the real valued and
symmetric operators ${\cal L}(F)\, v_n = Ag'(u_n) v_n - (v_{n+1}-2
v_n+v_{n-1})$, in spaces of sequences decaying exponentially at infinity.
Their spectra are discrete and real (these operators are compact), and we
would like to make sure that they are not empty. Since we are interested
mainly in the smallest eigenvalue, we shall use its variational
characterization, prove that this eigenvalue exists and characterize its
dependence on the parameters $A$ and $F$. We shall also describe finite
dimensional approximations of eigenvalues and eigenfunctions.

Let us first look for necessary conditions for $\lambda(A,F)$ to exist. Let
$\lambda \in \RR$ be an eigenvalue of ${\cal L}(F)$ with eigenfunction $V_n$.
Multiplying ${\cal L}(F)\, V_n - \lambda V_n$ by $V_n$ and summing over $n$,
we obtain
\begin{eqnarray}
0=\sum_n(V_{n+1}-V_n)^2 + [Ag'(u_n)-\lambda] V_n^2\quad\quad\quad\quad\quad
\nonumber\\
\geq \sum_n(V_{n+1}-V_n)^2 + [A\, \mbox{min}_n(g'(u_n)]-\lambda)\sum_n
V_n^2 .
\end{eqnarray}
Thus, $\lambda=\sum_n [(V_{n+1}-V_n)^2 + Ag'(u_n) V_n^2 ]/\sum V_n^{2} > A
\,\mbox{min}_n g'(u_n)$. This inequality implies that $\lambda$ is positive if
$u_n$ does not take on values in the region where $g'$ is negative, which
occurs for large enough $A>0$. In general, we can only say that $\lambda> A
g'(U_2)$ for $g'$ attains its minimum value in $[U_1,U_3]$ at $U_2$.
Similarly,
\begin{eqnarray}
0= \sum_n(V_{n+1}-V_n)^2 + [Ag'(u_n)-\lambda] V_n^2\nonumber\\
\leq [4+ A\, \mbox{max}_n(g'(u_n))-\lambda]\,\sum_n V_n^2
\end{eqnarray}
Therefore, $\lambda < A g'(U_1)+4=A g'(U_3)+4$, for $g'(u)$ attains its
maximum value in $[U_1,U_3]$ at the end points, $u=U_1$ and $u=U_3$.

The smallest eigenvalue $\lambda$ is given by the Rayleigh formula
\begin{equation}
\lambda = \mbox{Min}\, {\sum_n(w_{n+1}-w_n)^2 + Ag'(u_n) w_n^2 \over
\sum_n w_n^2}\,, \label{inf}
\end{equation}
where the infimum is taken over our space of exponentially decaying
functions. That Minimum is attained at an eigenfunction $V_n$ solving
(\ref{lin0}). Now, (\ref{lin0}) may have solutions decaying at $\pm \infty$
only if the difference equation
\begin{equation}
V_{n+1}+(-2-Ag'(U_1)+\lambda)\, V_n +V_{n-1} = 0 \label{diff}
\end{equation}
has solutions of the form $r^n$ with $r<1$. This happens when $(-2-A g'(U_1) +
\lambda)^2>4$. Thus either $\lambda>4+Ag'(U_1)>0$ (excluded above) or
$\lambda < Ag'(U_1)$. We conclude that $\lambda < A g'(U_1)$ is a necessary
condition to attain the Minimum (\ref{inf}) at a positive eigenfunction
decaying exponentially at infinity.

We now establish sufficient conditions for the Minimum (\ref{inf}) to exist.

{\bf Lemma A.1.- Conditions for the existence of positive decaying
eigenfunctions} {\it Let $F=0$, $A>0$, and let the nonlinearity $g$ satisfy
the hypotheses in Theorem A.1 Let $u_n$ be a stationary increasing solution
of (\ref{Fd}) such that $u_{-\infty}=U_1(0)$ and $u_{\infty}=U_3(0)$. Given an
exponentially decaying sequence $w=w_n$, we define
$$J(w)={\sum_n [(w_{n+1}-w_n)^2 + Ag'(u_n) w_n^2 ]\over \sum_n w_n^2}.$$
Let us suppose that there is a sequence $w_n$ such that $J(w_n)<A g'(U_1)$.
Then the infimum}
\begin{eqnarray} \lambda= \mbox{Inf}_{\sum_n [r_0^{-2|n|}v_n^2<\infty}\,
{\sum_n(v_{n+1}-v_n)^2 + Ag'(u_n) v_n^2 ]\over \sum_n v_n^2}
\label{lambda0}
\end{eqnarray}
{\it is attained at a positive function $V_n$ which decays as
$r(A,\lambda)^{|n|}$ at infinity, with
$0<r(A,\lambda)=[2+Ag'(U_1)-\lambda -\sqrt{(-2-A g'(U_1) +\lambda)^2 - 4}]/2 <
r_0 <1$. $\lambda=\lambda_1(A,0)$ and $v_n$ solves (\ref{lin0}).
}
\bigskip

\noindent 
{\it Remark A.1.} The value $0< r_0< 1$ is determined in the proof.
Note that $r(A,\lambda)$ is a decreasing function of $A$ but an
increasing function of $\lambda$.

\bigskip

\noindent 
{\it Remark A.2.} We have shown above that $Ag'(U_2)<\lambda < Ag'(U_1)$.
Thus the smallest eigenvalue shrinks to zero as $A\to 0$, although we do not
have a proof that it does so monotonically.

{\bf Proof:}
Clearly $J(w)$ is bounded from below by $A\, \mbox{min}_n(g'(u_n))$. We choose
$r_0=r(A,J(w_n))\in (0,1)$ and define $\|w\|_0=\sum r_0^{-2|n|}|w_n|^2$. Let
$w^m= w_{n}^m$, $m>0$ be a sequence minimizing $J(w)$: $\|w^m\|_0 <\infty$ and
$J(w^m) \rightarrow \lambda$ when $m \rightarrow \infty$. We replace $w^m$
with $v^m=v_{n}^m=w_{n}^m/ \|w^m\|_0$. Then, $\|v^m\|_0 =1$ and $J(v^m)
=J(w^m) \rightarrow \lambda$. $\|v^m\|_0 =1$ implies that, uniformly in $m$,
$|v_{n}^m|\leq r_0^{ |n|}$ and $\sum_{n>n(\epsilon)} |v_{n}^m|^2 <\epsilon$
for $n(\epsilon)$ large enough. Thus, a subsequence $v^m$ tends to some limit
$V=V_n$ such that $|V_{n}|\leq r_0^{|n|}$ and $\sum_n |v_{n}^m-V_n|^2$ tends
to zero as $m$ tends to infinity. Therefore, $J(v^m)\longrightarrow J(V)=
\lambda$ and the infimum is attained at the sequence $V=V_n$. Moreover,
$V=V_n$ satifies the Euler equation (\ref{lin0}) for the minimization
problem, which then implies that $V_n$ decays as stated in the Lemma. 

On the other hand, $J(|V_n|)\leq J(V_n)$ and we can choose nonnegative $V_n$.
But then $\lambda$ has to be the smallest eigenvalue $\lambda_1(A,0)$.
\fin

{\bf Lemma A.2.- Choice of the sequence $w_n$ with $J(w_n)< A g'(U_1)$.}
{\it Let $F=0$, $A>0$, and $u_n$ be as in Lemma A.1. Let $u(x)$ be the
solution of the boundary value problem $d^2u/dx^2 = g(u)$, with $u(-\infty) 
=U_1$, $u(\infty)=U_3$ such that $u(0)=U_2$. }

\begin{itemize}
\item {\it For sufficiently small $A<1$, we have}
 \begin{eqnarray}
|u_{n+1}-u_n|\, \mbox{max}_{[u_n,u_{n+1}]}\, |g''| < 2g'(U_1), \quad
\forall n,
\end{eqnarray}
{\it and $w_n=u_{n+1}-u_n$ satisfies $J(w_n)< A g'(U_1)$. An estimation of
the appropriate values of $A$ indicates that they should be smaller than}
$$A<\left( {2 g'(U_1)\over \mbox{max}_{[U_1,U_3]}
|g''|\,\mbox{max}_{\pRR} |du/dx|}\right)^2\,.$$
\item {\it For $A>2\, g'''(U_1)$, we can choose $w_n=0$ for $|n|>M\geq 0$,
$w_n= r^n$ for $n\geq 0$ and $w_n= r^{n+1}$ for $n< 0$.}
\end{itemize}  

{\bf Proof:}
The function $w_n=u_{n+1}-u_n$ is a solution of
$$w_{n+1}-2w_n+w_{n-1} = A{g(u_{n+1}) -g(u_n) \over u_{n+1}-u_n} w_n$$
that decays at infinity as $ r(A,0)^{|n|}$. Multiplying this equation by $w_n$
and adding over $n$, we obtain
\begin{eqnarray}
\sum_n\left((w_{n+1}-w_n)^2 + A{g(u_{n+1}) -g(u_n) \over u_{n+1}-u_n} w_n^2
\right) = 0.
\label{ecdifdis}
\end{eqnarray}
This result can be used to calculate $J(w_n)$:
\begin{eqnarray}
J(w_n)= A{\sum \left(g'(u_n)-{g(u_{n+1}) -g(u_n) \over u_{n+1}-u_n})\right)
w_n^2 \over \sum w_n^2} = -{A \over 2}\, {\sum g''(\xi_n) w_n^3 \over \sum
w_n^2}\,,
\end{eqnarray}
by the mean value theorem. Thus, $J(w_n)\leq  (A/2)$ max$_n\, (|u_{n+1}-u_n|$
max$_{[u_n,u_{n+1}]} |g''| )$. For sufficiently small $A$, $|u_{n+1}-u_n|
\leq C \sqrt{A}$, so that $J(w_n) < C A^{{3\over 2}}$ max$|g''|/2 < A
g'(U_1)$. More precisely, for small $A$, $w_n=u_{n+1}-u_n\simeq u((n+1)
\sqrt{A}) - u(n\sqrt{A})\simeq \sqrt{A} u'(\xi)$. Then $|w_n|\leq$ max
$|du/dx|\,\sqrt{A}$.

To prove the other case, we observe that $J(w_n)= (r-1)^2 + A \sum_{i=0}^{
\infty} g'(u_n) r^{2n}$ is smaller than $A g'(u_1)$ provided $(1-r)/(1+r) <
A\, g'''(U_1)/[2\, (1-(r(A,0) r)^2)]$ with $r(A,\lambda)$ defined as in Lemma
A.1. The last inequality holds if $A>2\, g'''(U_1)$. 
\fin
\bigskip

\noindent 
{\em Remark A.3}. For the FK nonlinearity, the first condition of
the Lemma holds for $A<0.9$, and the second condition for $A>2$. For
intermediate values, numerical simulations show that $w_n=u_{n+1}-u_n$
satisfies $J(w_n)< A g'(U_1)$.

{\bf Proposition A.1. - Finite dimensional approximations}
{\it  Let $u_n(A,F)$ be a stationary solution of (\ref{Fd}) under the
hypotheses in Theorem A.1 for $|F|\leq F_c(A)$. Let $\lambda_1(A,F)$ the
smallest eigenvalue of the operator ${\cal L}(F)$ (linearized about
$u_n(A,F)$) and $\lambda(A,F,N)$ be the smallest eigenvalues of the matrices
(\ref{red3}). Then, $\lambda(A,F,N)\rightarrow\lambda_1(A,F)$ as $N
\rightarrow \infty.$  As a consequence, if $V>0$ is an eigenfunction
associated to $\lambda_1(A,F)$ with $\sum_n V_n^2=1$ and
$V(N)>0$ are eigenvectors associated to $\lambda(A,F,N)$ such that $\sum_n
V_n(N)^2=1$, $V(N) \rightarrow V$ as $N \rightarrow \infty$.
}

{\bf Proof:} It follows from the Rayleigh characterizations for the
smallest eigenvalues:
\begin{equation}
\lambda_1(A,F) = \mbox{Min}_{\sum r^{-2|n|} w_n^2} {\sum_{-\infty}^{\infty}\,
[(w_{n+1}-w_n)^2 + Ag'(u_n(A,F)) w_n^2 ] \over \sum_{-\infty}^{\infty} w_n^2}
\end{equation}
\begin{equation}
\lambda(A,F,N) = \mbox{Min}\, {\sum_{-L}^{M}
[(v_{n+1}-v_n)^2 + Ag'(u_n(A,F)) v_n^2 ] \over
\sum_{-L}^{M} v_n^2}
\label{raleigh}
\end{equation}
Letting $w_n=v_n$ for $n=-L,\ldots, M$ and $w_n=0$ otherwise, we see that
$\lambda_1(A,F) \leq \lambda(A,F,N)$, $N=L+M+1$. Let now $w_n$ be an
eigenfunction for $\lambda_1(A,F)$ such that $\sum_{-\infty}^{\infty}
w_n^2=1$. Then,
$$\lambda(A,F,N)\leq  { \lambda_1(A,F) - \sum_{n<-L, n>M}
[(w_{n+1}-w_n)^2 + Ag'(u_n(A,F)) w_n^2 ]\over \sum_{-L}^{M} w_n^2}\,.$$
We conclude that $\lambda(A,F,N)\rightarrow\lambda_1(A,F)$ as $N \rightarrow
\infty$. This and the exponential decay of $V$ prove the convergence of the
eigenvectors.
\fin

\subsection{Proof of Theorem A.1}

The theorem will be proved in two steps and for simplicity, in the
particular case of periodic $g$. In this case, $U_3(F/A)-U_3(0) =
U_1(F/A)-U_1(0)$, which allows us to use symmetric sub and
supersolutions. Small modifications are required in the general case.

{\bf Step 1:} $F_c(A)>0$.

We use the existence of a positive eigenfunction $v_n$ associated
to a positive eigenvalue $\lambda_1(A,0)$ to construct stationary
supersolutions for (\ref{Fd}) when $F>0$ small.
The known solution $u_n$ provides a stationary subsolution.

We look for a supersolution of the form
\begin{eqnarray} w_n=u_n +(1+\delta)(U_1(F/A)-U_1(0))+\epsilon v_n
\label{stsup1}  \end{eqnarray}
with $\delta>0$ to be chosen and $\epsilon, F$ small to
be determined. Let us check  that
\begin{eqnarray} w_{n+1}-2w_n+w_{n-1} \leq g(w_n) -F,
\quad w_{\infty}>U_3(F/A), w_{-\infty}>U_1(F/A) \label{stsup2}
\end{eqnarray}
holds. The conditions at infinity are satisfied for any $\delta>0$.
Provided 
\begin{eqnarray} 
|(1+ \delta) (U_1(F/A)-U_1(0))|\leq k\epsilon, \label{cond1}
\end{eqnarray}
inequality (\ref{stsup2}) holds if
\begin{eqnarray} 
\epsilon\, (v_{n+1}-2v_n+v_{n-1}) < Ag'(u_n)\, [\epsilon v_n + (1+ \delta)
(U_1(F/A)-U_1(0))] -F + O(A\epsilon^2). \nonumber
\end{eqnarray}
Using (\ref{lin0}) we are left with
\begin{eqnarray} 
F < \epsilon\lambda_1(A,0) v_n + A (1+\delta) g'(u_n)
(U_1(F/A) -U_1(0)) \nonumber 
\end{eqnarray}
Now, $U_1(F/A)= g^{-1}({F\over A})$, the inverse being taken
near $U_1(0)$, in the region with $g'>0$. Using
$g^{-1}(x)\sim g^{-1}(x_0)+(g^{-1})'(x_0) (x-x_0)$ we obtain
\begin{eqnarray}
U_1(F/A)=g^{-1}(F/A)\sim U_1(A,0) + {F\over A\, g'(U_1(0))}\,.\label{cond2} 
\end{eqnarray} 
Thus, the condition for $w_n$ to be a supersolution is
\begin{eqnarray}
 F < \epsilon\lambda_1(A,0) v_n +   g'(u_n)
{1+\delta \over  g'(U_1(0))} F . \nonumber
\end{eqnarray}
Let $M$ be sufficiently large. We distinguish two different ranges of
indices $n$:
\begin{itemize}
\item For $|n|>M$, $g'(u_n)>0$ and $v_n\ll 1$. Then the right hand side of
the previous inequality is dominated by the second term. We choose $\delta$
large enough to ensure $F < g'(u_n) (1+\delta)\, F/g'(U_1(0))$, that is,
$1+\delta > g'(U_1(0))/ g'(u_M)$. 
\item For small $|n|$, $g'(u_n)<0$. The previous inequality is satisfied
provided we choose $F$ so small that 
\begin{eqnarray}
\left(1+ |g'(u_n)| {1+\delta \over g'(U_1(0))}\right)\, F <
\epsilon \lambda_1(A,0) v_n ,\nonumber
\end{eqnarray}
for a fixed value of $\delta$.
\end{itemize} 
With these choices, $w_n$ satisfies (\ref{stsup2}). Note that these choices
are compatible with condition (\ref{cond1}). Using (\ref{cond2}),
(\ref{cond1}) becomes $(1+\delta)\, F/[A\, g'(U_1(0))] < k\, \epsilon$. This
holds for small enough $F$. 

Let $F>0$ be small enough for a $w_n$ defined in (\ref{stsup1}) to be a
supersolution with $\delta,\epsilon$ adequately selected. Now, let $h_n(t)$
be a solution to (\ref{Fd}) for such $F>0$ with initial datum $h_n(0)$
satisfying $u_n< h_n(0) < w_n$. Then, $u_n < h_n(t) < w_n$, for all $t>0$.
Therefore, propagation is excluded and the solutions are pinned.

Stationary solutions $u_n(A,F)$ for such $F>0$ can be obtained as
long time limits of solutions $h_n(t)$ to (\ref{Fd}) when $h_n(0)$
is increasing, tends exponentially to $U_1(A,F)$ (resp.
$U_3(A,F))$ at $-\infty$ (resp. $\infty$) and $u_n< h_n(0) < w_n$.
We conclude that $F_c(A)>0$.

{\bf Step 2:} $\lambda_1(A,F)>0$ for $|F|<F_c(A)$ and
$\lambda_1(A,F_c(A))=0$

To fix ideas, we take $F>0$. The case $F<0$ follows by symmetry.
{}From Step 1, we know that $F_c(A)>0$ and there are stationary solutions
$u_n(A,F)$ of (\ref{Fd}) exist for $F>0$ small, that are increasing from
$U_1(F/A)$ to $U_3(F/A)$.

In an analogous way as we did for $F=0$  we get:
\begin{eqnarray}
\lambda_1(A,F) = \mbox{Min}_{\sum r_0^{-2|n|}w_n^2<\infty }
{\sum_n [(w_{n+1}-w_n)^2 + Ag'(u_n(A,F)) w_n^2]\over \sum w_n^2}\,.
\label{lambdaF}
\end{eqnarray}
This formula defines $\lambda_1(A,F)$ as a continuous function
of $F$. That $\lambda_1(A,0)>0$ implies $\lambda_1(A,F)>0$ up to some
$F_c$ at which $\lambda_1(A,F_c)=0$. As long as $\lambda_1(A,F_1)>0$, we
can obtain stationary solutions for $F>F_1$ (close to $F_1$), as done in
Step 1. This procedure cannot continue forever since such stationary
solutions do not longer exist for $F$ close to $A$: eventually $g(U)= F/A$
ceases to have three solutions and the stationary wave fronts cannot be
constructed. Thus, we must reach a value $F_c$ at which $\lambda_1(A,F_c)
= 0$.
 \fin

\vspace{.5cm}
{\bf Acknowledgments.}
The authors are indebted to J.M.\ Vega for a critical reading of the
manuscript and helpful comments and to V.\ Hakim for pointing out to them the
relevance of Ref.\ \cite{cah60}. A.C.\ thanks S.P.\ Hastings and J.B.\
McLeod  for fruitful discussions.

%\end{multicols}

\end{document}